\documentclass[journal,doublecolumn]{IEEEtran}
\usepackage[T1]{fontenc}

\usepackage{amsfonts}
\usepackage{amsmath}
\usepackage{booktabs}
\usepackage{footnote}
\usepackage{mathrsfs}  
\usepackage{hyperref}

\usepackage{float}
\usepackage{xcolor}
\usepackage{changes}
\definechangesauthor[name={R1}, color=orange]{R1}
\definechangesauthor[name={R2}, color=orange]{R2}
\definechangesauthor[name={R3}, color=orange]{R3}
\usepackage{multirow}
\usepackage{threeparttable}
\usepackage{amssymb}
\usepackage{stfloats}
\usepackage{graphicx}
\usepackage{placeins}
\usepackage{amsthm}
\usepackage{bbm}
\usepackage{subfigure}
\usepackage{array}
\usepackage[ruled,vlined]{algorithm2e}
\usepackage{bm}
\usepackage[square, comma, sort&compress, numbers]{natbib}
\usepackage{fancynum}
\definecolor{light-gray}{gray}{0.92}
\usepackage{colortbl}
\newtheorem{theorem}{Theorem}

\newtheorem{lemma}{Lemma}
\newtheorem{prop}{Proposition}
\newtheorem{defn}{Definition}
\newtheorem{remark}{Remark}
\newtheorem{assum}{Assumption}

\usepackage{mdframed}
\usepackage{nomencl}
\makenomenclature

\begin{document}
%
\title{Riemannian Low-Rank Model Compression for  Federated Learning {with Over-the-Air Aggregation}}
\author{Ye~Xue,~\IEEEmembership{Member,~IEEE,}
	and Vincent~LAU,~\IEEEmembership{Fellow,~IEEE}
	\thanks{This work was supported in part by the Research Grants
Council (RGC), Hong Kong, under Project 16207221, and in part by the
Huawei Technologies. (Corresponding author: Ye Xue) \\
   Y. Xue is with  Shenzhen Research Institute of Big Data, Shenzhen,  China (e-mail: yokoxue@sribd.cn). \\
   V. Lau is with the Department of Electronic and Computer Engineering, Hong Kong University of Science and Technology, Hong Kong, China.}}%




\maketitle

\begin{abstract}
Low-rank model compression is a widely used technique for reducing the computational load when training machine learning models. However, existing methods often rely on relaxing the low-rank constraint of the model weights using a regularized nuclear norm penalty, which requires an appropriate hyperparameter that can be difficult to determine in practice. Furthermore, existing compression techniques are not directly applicable to efficient over-the-air (OTA) aggregation in federated learning (FL) systems for distributed Internet-of-Things (IoT) scenarios. In this paper, we propose a novel manifold optimization formulation for low-rank model compression in FL that does not relax the low-rank constraint. Our optimization is conducted directly over the low-rank manifold, guaranteeing that the model is exactly low-rank. We also introduce a consensus penalty in the optimization formulation to support OTA aggregation. Based on our optimization formulation, we propose an alternating Riemannian optimization algorithm with a precoder that enables efficient OTA aggregation of low-rank local models without sacrificing training performance. Additionally, we provide convergence analysis in terms of key system parameters and conduct extensive experiments with real-world datasets to demonstrate the effectiveness of our proposed Riemannian low-rank model compression scheme compared to various state-of-the-art baselines.
\end{abstract}


%
\IEEEpeerreviewmaketitle

\section{Introduction}\label{sec:intro}
{
\IEEEPARstart{T}{he}  past decade has witnessed a revolution in the data processing paradigm due to the proliferation of machine learning techniques. However, many successful machine-learning approaches rely on training models with millions or even billions of parameters. For example, the popular convolutional model VGG16 \cite{Simonyan15} has about $10^8$ parameters, and the powerful language model Generative Pre-trained Transformer 3 (GPT-3) \cite{NEURIPS2020_1457c0d6}, has a capacity of $175$ billion parameters. This demand for computational resources and data makes the deployment of AI to mobile phones or smart devices challenging. To reduce computational loading, model compression \cite{gong2014compressing,wu2016quantized,frankle2020linear,han2015learning,lebedev2016fast,kumar2021pruning,vogels2019powersgd,cho2019gradzip,lowrankrelax,yuan2021federated} explores the low-dimensional structures of the learning models to efficiently compress the high-dimensional model weights.

Model compression techniques can be categorized along many lines, including parameter  quantization \cite{gong2014compressing,wu2016quantized}, network pruning \cite{frankle2020linear,han2015learning},  $\ell_1$ regularization \cite{lebedev2016fast,kumar2021pruning},  low-rank model compression \cite{vogels2019powersgd,cho2019gradzip,lowrankrelax,NEURIPS2020_a376802c,yuan2021federated}, etc. Parameter quantization compresses the original network by reducing the number of bits required to represent each weight. However, this method can only achieve limited compression. Network pruning aims to remove redundant, non-informative weights in the models. A typical procedure of network pruning requires training a large, overparameterized model first, then pruning the trained large model into a sparse model according to a certain criterion. Although it can reduce the computation load during inference, the requirement of the large pre-trained model costs huge computational resources during training \cite{frankle2020linear}. The $\ell_1$ regularization method \cite{kumar2021pruning} imposes the $\ell_1$ regularization of model weights to obtain a sparse model. Unlike classical network pruning, this method can support training from scratch, avoiding the computation of a pre-trained model. However, the resulting weight matrices of both network pruning and $\ell_1$  regularization will be randomly sparse, making the datapath design very challenging \cite{cheng2017survey}.}  Low-rank model compression aims to learn low-rank model weights and has the efficiency advantage in both compression and computation. Specifically, an  $M\times N$ weight matrix $\bm{\Theta}$ with rank $R$  is represented by the product of two smaller matrices $\bm{U}$ and $\bm{V}$ with sizes $M\times R$ and $R\times N$.  Usually, $M$ and $N$ are very large and $R$ can be rather small.  Hence, the low-rank method can achieve a high compression ratio. Furthermore, the computational loading during inferencing can be dramatically reduced. For example, the computation complexity of the product between the uncompressed weight matrix $\bm{\Theta}\in\mathbb{R}^{M\times N}$ and an input data vector $\bm{x}\in\mathbb{R}^{N}$ is $\mathcal{O}(MN)$. By low-rank compression, this product can be computed via $\bm{\Theta}\bm{x}=\bm{U}\bm{V}\bm{x}$,  which reduces the complexity to  $\mathcal{O}(R(M+N))$.   Moreover,  since the computation of $\bm{U}\bm{V}\bm{x}$ only involves general matrix-vector product, no specific hardware datapath design is required. Therefore it can be easily implemented in a general hardware environment \cite{yang2020learning}.

Due to the aforementioned benefits, many works have focused on low-rank model compression. In \cite{vogels2019powersgd,cho2019gradzip}, the authors proposed to approximate the unconstrained local model with low-rank matrices. Such a brute-force solution reduces the model size at the cost of large approximation errors and degraded training performance. Other works resort to holistic optimization methods based on the low-rank constraint for the model \cite{lowrankrelax,yuan2021federated}. Since the low-rank constraint is non-convex, which is NP-hard from a traditional optimization perspective, researchers adopt convex relaxation in the form of a regularized nuclear-norm penalty to avoid non-convexity  \cite{lowrankrelax,yuan2021federated}. The resultant optimization problem at the $k$-th local device is as follows: 
\begin{equation}
\begin{aligned}
\min_{\bm{\Theta}_{k}\in \mathbb{R}^{M\times N}}\quad & \frac{1}{|\mathcal{D}_k|}\sum_{l_k\in \mathcal{D}_k}f(\bm{\Theta}_k;\bm{\zeta}_{l_k})+\lambda \|\bm{\Theta}_k\|_*,
\end{aligned}\label{prob:nuclear}
\end{equation}
where $f(\cdot)$ is the optimization loss in terms of  the optimization weights, $\bm{\Theta}_{k}$, and the $l_k$-th data sample, $\bm{\zeta}_{l_k}$, from the local data set, $\mathcal{D}_k$.  $\lambda$ is the regularization parameter, and $\|\cdot\|_{*}$ is the nuclear norm. However, this convex relaxation technique also raises various problems. First, the rank of the local model is highly sensitive to the choice of the regularization parameter $\lambda$. An appropriate choice of the regularization parameter $\lambda$ depends on the statistics of the data, which is usually not available in practice. Second, although the nuclear norm is convex, it is non-smooth. In modern machine learning tasks, such as deep neural network (DNN) training, $f(\bm{\Theta_k};\bm{\zeta}_{l_k})$ in (\ref{prob:nuclear})  is usually non-convex in $\bm{\Theta}_{k}$.  Therefore, the overall optimization is non-convex and non-smooth, which is quite challenging to solve by low-complexity first-order methods due to the irregular optimization landscape \cite{yuan2021federated}. {Furthermore, to
analyze the convergence property,  nontrivial assumptions (such as Kurdyka-Łojasiewicz
conditions) are usually required \cite{attouch2013convergence}}.    

{
In all the aforementioned schemes, model compression has been considered in centralized settings where the model and the data reside in the cloud. To embrace future scenarios where the data sets reside distributively at the IoT devices and protect privacy, federated learning (FL) has been proposed \cite{10.1145/3298981,mcmahan2017communication}. The FL settings will pose additional technical challenges to model compression. The aforementioned solutions cannot be directly applied because the communication resources during the training have been ignored in conventional model compression solutions. For example,  over-the-air (OTA) aggregation \cite{sery2020analog,9042352,9050465} has been proposed to reduce the radio resource required for FL training with a large number of IoT devices \cite{yang2020federated}. Exploiting the free aggregation of the transmitted symbols over the wireless channels, multiple IoT devices can share a common radio resource block for gradient aggregation or weight aggregation in the uploading process. However, conventional $\ell_1$ regularization or low-rank model compression solutions are not compatible with the OTA  aggregation in FL because the sparsity or low-rank properties of the model weight matrix may be destroyed after the direct model aggregation and therefore decelerate the training.}

In this work, we propose a novel manifold optimization formulation with a consensus penalty to obtain  exact low-rank local models, which will be compatible with the OTA aggregation  during FL training.   An efficient alternating Riemannian optimization algorithm is then proposed with a convergence  guarantee. We summarize the key contributions below.
\begin{itemize}
	\item {\bf Low-Rank Manifold  Optimization Formulation for Model Compression in FL:} We propose a novel optimization  formulation over the low-rank manifold for FL systems without convex relaxation.  As such, the low-rank structure of the local models {is always maintained during the training.} In addition, the formulation also incorporates  a novel consensus penalty term that can enforce the local models {not to deviate} from each other. This key feature enables the global model to be still low-rank  after OTA  aggregation in FL training and will not influence the training trajectory.
	\item{\bf Riemannian Stochastic Algorithm with Over-the-Air Aggregation:}
To solve the proposed low-rank manifold  optimization problem, we exploit the benign Riemannian property of the low-rank manifold to propose a novel alternating Riemannian algorithm with OTA aggregation, in which a random linear coding (RLC)-based compression precoder is introduced to  mitigate the cross term issue during the OTA aggregation.  {In addition, we show that under some mild conditions, the proposed solution will converge to a Karush–Kuhn–Tucker (KKT) point.}
	\item {\bf Efficient Implementation on Real-World  Data-Set:}
We provide extensive experiments which show that the proposed scheme can achieve a better compression performance and fast CPU time  under very broad conditions compared to various state-of-the-art
baselines, including $\ell_1$ regularization and network pruning.
\end{itemize}
The rest of the paper is organized as follows. Section \ref{sec:sys}
introduces the system models, including the federated training model and the wireless communication model.  Section \ref{sec:prob}  presents the proposed manifold optimization problem formulation. Based on the formulation,  Section \ref{sec:alg} and \ref{sec:conv} demonstrate the proposed Riemannian algorithm  and the corresponding convergence analysis, respectively.  The numerical experiments are 
given in Section \ref{sec:exp} and Section \ref{sec:con} summarizes the work.

\emph{Notations}: 
In this paper, lowercase and uppercase boldface  letters stand for column vectors and matrices, respectively.   The operations $(\centerdot)^T$ and $\mathbb{E}[\centerdot]$ denote the operations of transpose and expectation, respectively.  We use $\langle \bm{X},\bm{Y}\rangle$ to denote the general inner product of $\bm{X}$ and $\bm{Y}$. $||\centerdot||_{p}$ is the  $\ell_{p}$-norm of a vector or the induced   $\ell_{p}$-norm  of a matrix and $||\centerdot||_{F}$ is the Frobenius norm of a matrix.$\sigma_{max}(\mathbf{A})$ is the largest singular value of matrix $\mathbf{A}$.   $[K]$ represents the set of integers $\{1,2,\ldots,K\}$. Set $\{\bm{\Theta}_k,\forall k\in[K]\}$ is is abbreviated as $\{\bm{\Theta}_k\}_{k=1}^K$. $\mathcal{ST}_{N}^{R}$  is the Stiefel manifold of $N\times R$ real orthonormal matrices. For a set $\mathcal{S}$, $|\mathcal{S}|$ denotes its cardinality. $\mathcal{P}_{\mathcal{S}}(\cdot)$ is a projection operator onto  set $\mathcal{S}$. $\lceil\centerdot \rceil$, $\lfloor\centerdot\rfloor$, and $|\centerdot|$ denote the ceiling operator,  floor operator, and taking the abstract
value.    $\angle{h}$ is the phase of the complex value $h$. $\nabla_{\bm{\Theta}}$ and $grad_{\bm{\Theta}}$ are the Euclidean gradient and the Riemannian gradient with respect to $\bm{\Theta}$, respectively.  $DF (\bm{A})[\bm{\Lambda}]$ is called the directional derivative of $F$ at $\bm{A}$ 
along direction $\bm{\Lambda}$.

\section{System Model}\label{sec:sys}
In this section, we review the basic system models for  FL and  wireless communication. In addition, we will discuss why  the conventional model compression methods are  incompatible   with the over-the-air aggregation when implementing FL in wireless systems.
\subsection{Federated Learning Model}\label{subsec:ft}
We consider an FL  system where $K$ IoT devices collaboratively
learn a machine-learning model with the help of a central server. Each IoT device $k$ has a local dataset $\mathcal{D}_k$ with $|\mathcal{D}_k|=D$.
The target of the system is to train a  model  {with trainable weight matrix} {$\bm{\Theta}\in \mathbb{R}^{M\times N}$}  by collaboratively optimizing the following problem:
\begin{equation}
\begin{aligned}
\underset{\bm{\Theta}\in\mathbb{R}^{M\times N}}{\min}F(\bm{\Theta})&=\frac{1}{\sum_{k=1}^K|\mathcal{D}_k|}\sum_{k=1}^K\sum_{l=1}^{|\mathcal{D}_k|}f(\bm{\Theta},\bm{\zeta}_{k,l})\\
&=\frac{1}{K}\sum_{k=1}^K\frac{1}{D}\sum_{l=1}^Df(\bm{\Theta},\bm{\zeta}_{k,l}),
\end{aligned}
\label{eq:sys}
\end{equation} {using} mini-batch stochastic gradient descent (MSGD). $\bm{\zeta}_{k,l}$ in Eq.(\ref{eq:sys}) denotes the $l$-th data sample from the local dataset $\mathcal{D}_k$ of the $k$-th device. A standard implementation of MSGD in the FL system  is FedAvg \cite{mcmahan2017communication}, which    usually contains the following  steps:
\begin{itemize}
	\item {\bf Step 1: Server Broadcasts the Global Model Weights (Downlink Transmission)}:  { At the $t$-th training iteration, the   updated weight matrix of the model,  $\bm{\Theta}^{(t)}\in\mathbb{R}^{M\times N}$,} are broadcast to each  device from the server.
	\item {\bf Step 2: Devices Compute Local Model Weights}: Each device $k$ updates the local model weights $\bm{\Theta}^{(t+1)}_k$ via
	 \begin{equation}\label{cal:grad}
	\bm{\Theta}^{(t+1)}_k =\bm{\Theta}^{(t)}-\eta_k^{(t)}\centerdot \frac{1}{|\mathcal{D}^{(t)}_k|} \nabla_{\bm{\Theta}} \sum_{\bm{\zeta}_{k,l}\in\mathcal{D}^{(t)}_k}f(\bm{\Theta}^{(t)},\bm{\zeta}_{k,l}),
	\end{equation}
	where $\mathcal{D}^{(t)}_k\subseteq\mathcal{D}_k$ denotes the uniformly random selected mini-batch of data used to calculate the stochastic gradient, $\frac{1}{|\mathcal{D}^{(t)}_k|} \nabla \sum_{\bm{\zeta}_{k,l}\in\mathcal{D}^{(t)}_k}f(\bm{\Theta}^{(t)},\bm{\zeta}_{k,l})$, at the $t$-th iteration. 	$\eta_k^{(t)} \in \mathbb{R}$ is the local learning rate.  Note that (\ref{cal:grad}) shows the single round of the local MSGD without loss of generality and the device can also update the local weights after multiple rounds of local MSGD.
	\item {\bf Step 3: Devices Upload Local Model Weights (Uplink Transmission)}: Each device  $k$ uploads its local model weights{ $\bm{\Theta}^{(t+1)}_k\in \mathbb{R}^{M\times N}$} to the central server for further aggregation.
	\item {\bf Step 4: Server Aggregates the Local  Model Weights}: The server collects an aggregate of the device updates to obtain the new global model, { ${\bm{\Theta}}^{(t+1)}\in \mathbb{R}^{M\times N}$}, via
		 \begin{equation}
		 \begin{aligned}
		 \bm{\Theta}^{(t+1)}&	=\frac{1}{K}\sum_{k=1}^K \bm{\Theta}^{(t+1)}_k.
		 		 \end{aligned}
		 \end{equation}
		 Then the server broadcasts the updated global model to each local device.
\end{itemize}
\subsection{Communication Model}
When implementing the aforementioned FL training procedures in wireless communication systems, the uplink transmission of the local weights from the devices  becomes a primary bottleneck because massive
devices will be involved  but the radio resource is limited. {The information in FL at the server is the superposition of the individual weights or gradients transmitted by the IoT devices. Using conventional multiple access schemes, the communication and computation are separated in the sense that the server has to first estimate the individual terms transmitted by each IoT device, and then the aggregation of terms is computed within the server. Such a conventional approach will incur huge communication overheads especially when a huge number of IoT devices are involved. On the other hand, OTA aggregation is proposed \cite{sery2020analog,9042352,9050465} to exploit the additive nature and therefore the free aggregation of signals in the wireless multiple access channel (MAC). Specifically, the server can estimate the required quantity from the received signals directly without first estimating the individually transmitted terms. As such, the IoT devices can share a common radio resource block to upload the gradients or local weights to the server.}  The signal received at the server in the $t$-th training iteration is denoted as 
 \begin{equation}
 \begin{aligned}
\bm{Y}^{(t)}=\sum^K_{k=1}h_{k}^{(t)}p_{k}^{(t)}\bm{S}_{k}^{(t)} + \bm{Z}^{(t)},
 \end{aligned}
 \label{eq:tx}
 \end{equation}
 where { $\bm{S}_{k}^{(t)}\in\mathbb{R}^{M_{S}\times N_{S}}$ denotes  the transmitted signal matrix from device $k$, and $h_{k}^{(t)} \in\mathbb{C}$ represents the block channel coefficient {from the $k$-th device to the server},  which  follows Rayleigh fading in an  i.i.d.  manner over the indexes of $k$, $t$, i.e., $h_{k}^{(t)}\sim_{i.i.d}\mathcal{CN}(0,1)$.  {We consider a time-division duplex  system where  uplink-downlink channel reciprocity exists, and hence devices can estimate their  channel state information (CSI) through the pilots broadcast.}  $p_{k}^{(t)}\in\mathbb{C}$ is the transmission coefficient, and $\bm{Z}^{(t)}\in\mathbb{R}^{M_{S}\times N_{S}}$ denotes the i.i.d. additive white
 Gaussian noise (AWGN), i.e., each element in the noise matrix $\bm{Z}^{(t)}$ follows $\mathcal{CN}(0, \sigma_z^2)$ in an i.i.d manner for all $i$ and $t$.}

 {For the downlink broadcast from the central server to the local devices, we assume the transmission is error-free since the server has higher transmit power with forward error correction  compared to the uplink access  \cite{sery2020analog,9042352}.}

\begin{figure}[htbp]
	\centering
			\includegraphics[width=0.3\textwidth]{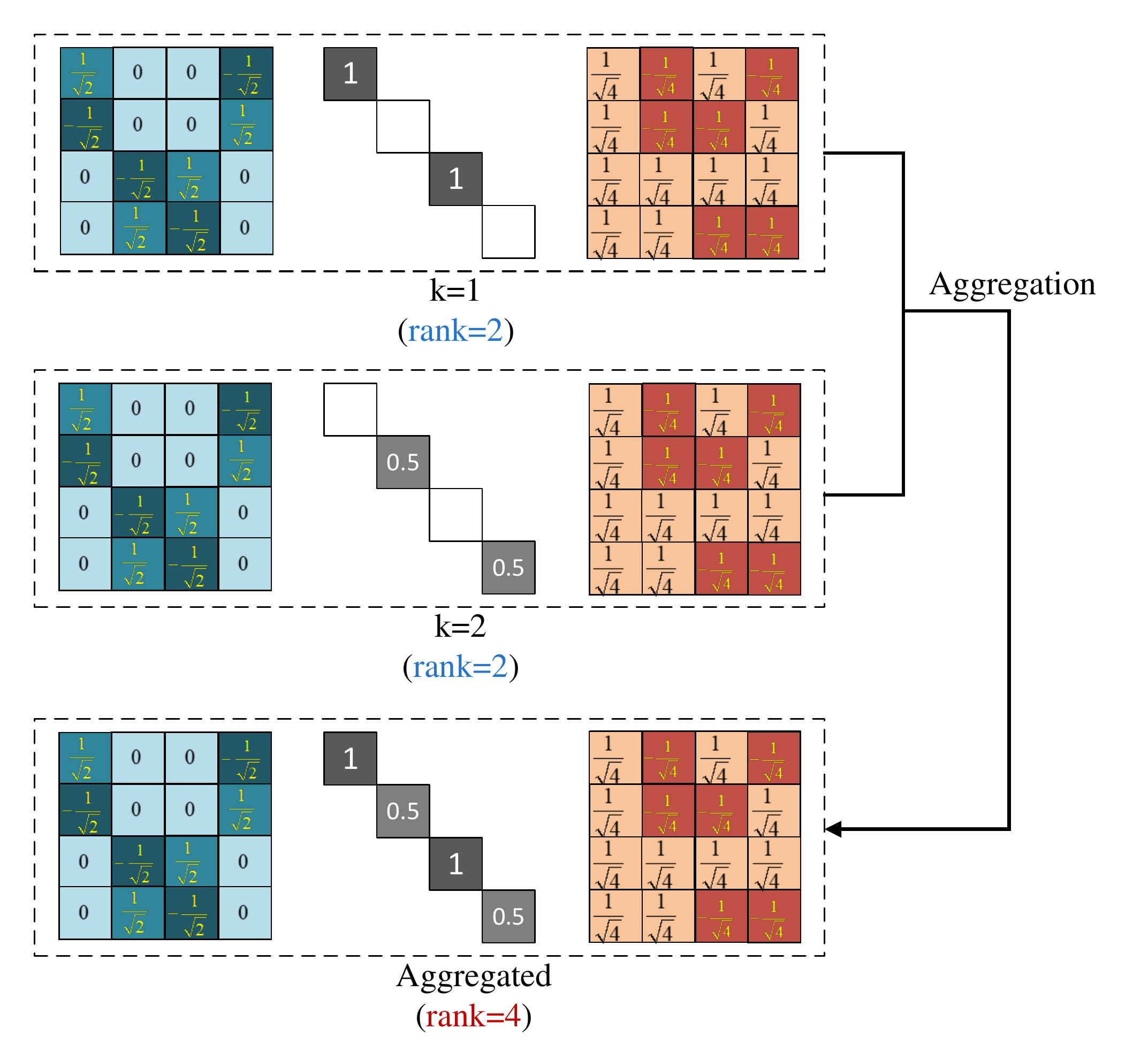}\label{fig:otaissuelr}
	\caption{Illustration of the destruction of the low-dimensional structure of the conventional model compression after OTA aggregation in FL. Low-rank model compression: each local model is of rank $2$, but the aggregated model is full rank with rank $4$.} \label{fig:otaissue}
\end{figure}

{\subsection{Incompatibility of the Conventional Model Compression to the Over-the-Air Aggregation in Federated Learning}\label{Sec:incomp}}
Though the above FL system with OTA aggregation can effectively protect data privacy and improve spectrum efficiency, it is incompatible to  low-rank model compression. In conventional model compression methods, the local devices will produce models with  low rankness  {on the local weight matrices.} However, the desired structure will be destroyed after OTA aggregation of these local models in FL. {      This is because  the low-rank pattern learned in the
trained local model at each device may diverge a lot from the low-rank pattern from other devices due to different local data sets.
Thus, the OTA aggregation at the server will destroy the low-rank structure as can be seen in Fig. \ref{fig:otaissue}. In this case, during
each iteration of local training, the device will need to induce the
structure pattern into the model from scratch, which can be quite
inefficient. In addition, each device will have to download an unstructured model
which prevents leveraging the structured pattern to save the communication
cost in the downlink.}

\section{Problem Formulation}\label{sec:prob}

 As mentioned in the previous sections, traditional low-rank compression formulation has the following drawbacks that prevent it to be efficiently applied in FL systems:
 \begin{enumerate}
     \item Due to the sensitivity of the regularization parameter, the model may not exactly satisfy the low-rank requirement.\label{issue:1}
     \item Due to the non-convexity and non-smoothness, the problem is hard to optimize.\label{issue:2}
     \item Due to the deviation of the local model weights, the scheme is incompatible with the OTA aggregation.\label{issue:3}
 \end{enumerate}
 In this section, we will propose a novel formulation that can tackle these issues.

To guarantee that the model strictly  satisfies the low-rank requirement, we propose to constrain the local training model  to the low-rank manifold $\mathcal{M}_R$ and to optimize the following problem in the FL system
\begin{equation}
\begin{aligned}
\mathscr{P}1:\quad\underset{\bm{\Theta}_{0}\in\mathcal{M}_R}{\min}F(\bm{\Theta}_0)&=\frac{1}{K}\sum_{k=1}^K\frac{1}{D}\sum_{l=1}^Df(\bm{\Theta}_{0},\bm{\zeta}_{k,l}),
\end{aligned}
\label{eq:sysP1}
\end{equation} 
{     where 
\begin{equation}
\begin{aligned}
\mathcal{M}_R=\{\bm{\Gamma}\in\mathbb{R}^{M\times N}: rank(\bm{\Gamma})=R\}
\end{aligned}
\label{eq:rankm}
\end{equation}
is the set of ${M\times N}$ real matrices with rank $R$. This set is defined as the rank-$R$ manifold. }

\begin{figure}[H]
	\centering
	\includegraphics[width=0.6\linewidth]{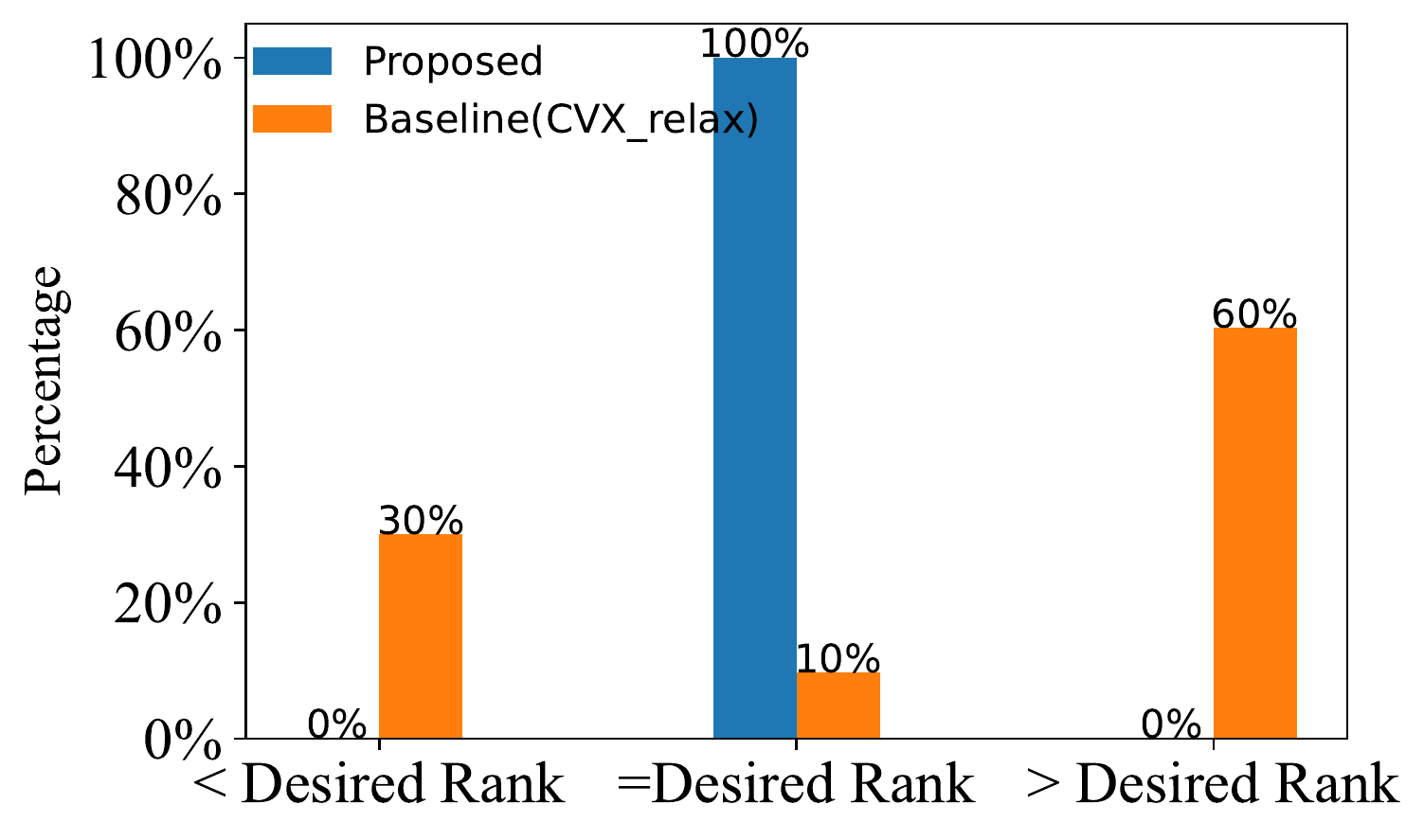}
	\caption{Rank of the local model weights produced by different methods. Experiments are carried out on the local training for low-rank model compression with weight matrix $\bm{\Theta}_{0}\in\mathbb{R}^{1000\times 2000}$ and the desired rank $40$. {$Proposed$} denotes the results of proposed optimization over the low-rank manifold,  {$CVX_{relax}$} denotes the nuclear-norm relaxation solved by the proximal algorithm \cite{davenport2016overview} with $\lambda=1500$. {     The rank constraint can be considered as  the $l_0$-norm  constraint of the singular value of the matrix while the nuclear norm  is   the $l_1$-norm  of the singular value of the matrix. The proposed method directly optimizes the model in the fix-rank manifold, which preserves the original constraint.    The nuclear-norm relaxation method   relaxes  the rank constraint \cite{cai2010singular} to optimize in a larger space than the original constraint set, which will produce some  solutions that violate the rank constraint.}}
	\label{fig:vRANK}
\end{figure}

As shown in Fig. \ref{fig:vRANK}, the model produced by solving Problem (\ref{eq:rankm}) satisfies the desired rank while the  nuclear-norm regularized formulation may produce some instances that violate the rank constraint.
Another advantage of Problem (\ref{eq:sysP1}) is that the rank-$R$ manifold $\mathcal{M}_R$ is a smooth  Riemannian manifold because of the following proposition:
\begin{prop}[Smoothness of the Low-Rank Manifold]\label{prop:smooth}
$\mathcal{M}_R$ is a $C^{\infty}$ smooth embedded submanifold of   $\mathbb{R}^{M\times N}$  of dimension
\begin{equation}
dim(\mathcal{M}_R)=MN-(M-R)(N-R)=(M+N-R)R
\end{equation}
\end{prop}
\begin{proof}
For any matrix $\bm{A}\in\mathbb{R}^{M\times N}$, we have
\begin{equation}
\begin{aligned}
\bm{A}=\left[\begin{array}{c c c} 
   {\bm A}_{1,1} & {\bm A}_{1,2} \\ 
   {\bm A}_{2,1} & {\bm A}_{2,2}
\end{array} \right]
\end{aligned}
\end{equation}
 and $\bm{A}\in\mathcal{M}_R$ if and only if ${\bm A}_{2,2}-{\bm A}_{1,2}{\bm A}_{1,1}^{-1}{\bm A}_{2,1}=\bm{0}_{(M-R)\times (N-R)}$ by the property of the Schur complement \cite{zhang2006schur}.
 Then, we define a map $F: \mathbb{R}^{M\times N}\to  \mathbb{R}^{(M-R)\times (N-R)}$ as $F(\bm{A}) = {\bm A}_{2,2}-{\bm A}_{1,2}{\bm A}_{1,1}^{-1}{\bm A}_{2,1}$. Therefore, we have $F(\bm{0})^{-1}\in\mathcal{M}_R$. Then,  at each point $\bm{A}\in\mathcal{M}_R$, for all $\hat{\bm{\Lambda}}\in \mathbb{R}^{(M-R)\times (N-R)}$, there exists $\bm{\Lambda}\in \mathbb{R}^{M\times N}$ such that $DF (\bm{A})[\bm{\Lambda}]= \hat{\bm{\Lambda}}$, with $\bm{\Lambda}=\left[\begin{array}{c c c} 
   {\bm 0} & {\bm 0} \\ 
   {\bm 0} & \hat{\bm{\Lambda}}
\end{array} \right]$. Hence the $F$ is a submersion at each point $\bm{A}\in\mathcal{M}_R$. By the {\em submersion theorem} \cite [Proposition 3.3.3]{absil2009optimization}, we finish the proof.
\end{proof}
The smoothness of $\mathcal{M}_R$ guarantees the differentiability, hence, enabling efficient and low-complexity first-order Riemannian optimization algorithms. {However}, optimizing (\ref{eq:sysP1}) by solving its local copy on each device and then aggregating the local results by the server will make the FL training incompatible with the OTA aggregation since there will be a deviation of the local model weights as stated in Issue \ref{issue:3}. To address this issue, we propose a decomposed optimization formulation  with a consensus penalty and a Riemannian optimization algorithm with OTA aggregation in the following section.

\section{Decomposed Riemannian-based Alternating Optimization with OTA Aggregation}\label{sec:alg}
\subsection{Decomposed Optimization Formulation with Consensus Penalty}
First, to decompose Problem $\mathscr{P}1$ in (\ref{eq:sysP1}) for each local device, we introduce the auxiliary variables $\{\bm{\Theta}_k\in\mathcal{M}_R\}_{k=1}^K$ to represent the local optimization weights and equivalently transform  $\mathscr{P}1$  into  
\begin{equation}
\begin{aligned}
\mathscr{P}2:\underset{\{\bm{\Theta}_k\in\mathcal{M}_R\}_{k=1}^K, \bm{\Theta}_0\in \mathcal{M}_R}{\min} \quad &\frac{1}{K}\sum_{k=1}^K\frac{1}{D}\sum_{l=1}^Df(\bm{\Theta}_k,\bm{\zeta}_{k,l})\\
\textrm{s.t.}&\quad \bm{\Theta}_k=\bm{\Theta}_0, \quad k=1,\ldots,K.
\end{aligned}
\label{eq:sysiter}
\end{equation}
{The equality constraints on $\bm{\Theta}_k, \forall k \in[K]$ creates coupling to the local subproblems.} {To} enable efficient distributed algorithms, we relax Problem $\mathscr{P}2$  into the following formulation by transforming the equality constraints into a {\em consensus penalty} as
\begin{equation}
\begin{aligned}
\mathscr{P}3:\underset{\{\bm{\Theta}_k\in\mathcal{M}_R\}_{k=1}^K, \bm{\Theta}_0\in \mathcal{M}_R}{\min} \quad &\frac{1}{K}\sum_{k=1}^K\frac{1}{D}\sum_{l=1}^Df(\bm{\Theta}_k,\bm{\zeta}_{k,l})\\
&\quad+\frac{\mu}{2K}\sum_{k=1}^K\|\bm{\Theta}_0-\bm{\Theta}_{k}\|_F^2,\\
\end{aligned}
\label{eq:sys2}
\end{equation}
where $\mu>0$ is a penalty parameter. Ideally, the penalty term $\frac{\mu}{2K}\sum_{k=1}^K\|\bm{\Theta}_0-\bm{\Theta}_{k}\|_F^2$ achieves its minimum at $\bm{0}$ when the local model weights are identical, i.e., $\frac{\mu}{2K}\sum_{k=1}^K\|\bm{\Theta}_0-\bm{\Theta}_{k}\|_F^2\geq 0$, and the equality holds if and only if $\bm{\Theta}_{1}=\bm{\Theta}_{2}=\ldots=\bm{\Theta}_{K}=\bm{\Theta}_{0}$. Intuitively,  the value of $\mu$ should be made {large enough}, such that the penalty term will lead to a heavy cost for any  violation of the equality constraints  in $\mathscr{P}2$. In this case, the minimization of $\mathscr{P}3$ will yield a feasible solution to the original problem $\mathscr{P}2$. However, large values of $\mu$ create enormously flat valleys at the constraint boundaries. Flat valleys will often present  insurmountable convergence difficulties for all preferred search methods unless the algorithm has already produced a point extremely close to the target solution \cite{luenberger2016penalty}. Therefore, $\mathscr{P}3$ should be solved  with   monotonically increasing values of $\mu$.

\begin{remark}[Advantages of the Consensus Penalty]
The optimization  in existing  federated learning methods  only focuses on minimizing the task-driven loss $\frac{1}{D}\sum_{l=1}^Df(\bm{\Theta}_k,\bm{\zeta}_{k,l})$ on each local device but ignore seeking a consensus with other local model weights. This will cause incompatibility in the OTA aggregation. In the proposed Problem $\mathscr{P}3$, each local model weight $\bm{\Theta}_k$ will be influenced  by assessing a penalty equal
to the square of the violation of consensus. In this way, the local training will seek a balance between optimizing the task-driven loss and the consensus loss and the deviation among local model weights will be mitigated. Therefore, Problem $\mathscr{P}3$ {is}  compatible with  the OTA aggregation. {      Though  Problem $\mathscr{P}2$ is also compatible with  OTA aggregation due to the consensus constraint that prevents the aggregated model from destroying the  local low-rank structures, it is challenging to solve  Problem $\mathscr{P}2$  because the equality constraints create coupling effects on the local variables. Compared to $\mathscr{P}2$ with multiple equality constraints, $\mathscr{P}3$  only involves the manifold constraints after relaxing the equality constraints into the consensus penalty, which will enable efficient algorithms. Furthermore, as the penalty parameter $\mu$ in $\mathscr{P}3$  becomes larger, the consensus penalty has a greater weight along the training process. In this way, all the local models trained by optimizing Problem $\mathscr{P}3$  will converge to a consensus model, which is a feasible point in Problem $\mathscr{P}2$ as  we will show in Lemma \ref{lem:convconsen}.}
\end{remark}

\subsection{Alternating Optimization }
{$\mathscr{P}3$} can be decomposed into each local device and solved by low-complexity algorithms. Specifically, we adopt the alternating optimization (AO) framework to solve  $\mathscr{P}3$ in the FL system. At each iteration $t$, the following two steps will be executed in an alternate manner.
\begin{itemize}
    \item {\bf Step 1 (Update the Local Weights on Local Devices) :} Given $\bm{\Theta}_0^{(t)}$, the problem in terms of the variables $\{\bm{\Theta}_k\in\mathcal{M}_R\}_{k=1}^K$ can be decomposed into $K$ parallel subproblems  among $K$ devices. Specifically, the  model weight matrix on device $k$ is updated by
    \begin{equation}
\begin{aligned}
\bm{\Theta}_{k}^{(t+1)}=\underset{\bm{\Theta}_k\in\mathcal{M}_R}{\arg\min} \quad &\frac{1}{DK}\sum_{l=1}^Df(\bm{\Theta}_k,\bm{\zeta}_{k,l})\\
&\quad+\frac{\mu}{2K}\|\bm{\Theta}_{0}^{(t)}-\bm{\Theta}_{k}\|_F^2.\\
\end{aligned}
\label{eq:sys2local}
\end{equation}
\item {\bf Step 2 (Update the Global Weights on the Server):} After obtaining $\{\bm{\Theta}_k^{(t+1)}\}_{k=1}^K$ from (\ref{eq:sys2local}), the updated global weight matrix  is given by
    \begin{equation}
\begin{aligned}
\bm{\Theta}_{0}^{(t+1)}=\underset{\bm{\Theta}_{0}\in\mathcal{M}_R}{\arg\min} \quad &\frac{\mu}{2K}\sum_{k=1}^K\|\bm{\Theta}_{0}-\bm{\Theta}^{(t+1)}_{k}\|_F^2.\\
\end{aligned}
\label{eq:sys2global}
\end{equation}
\end{itemize}

Though the above updates naturally result in  distributed calculations among local devices, there are several challenges.
\begin{enumerate}
    \item There is no closed-form solution to the problem in (\ref{eq:sys2local}), due to the non-convexity of the loss and the manifold constraint.
    \item The solution to the problem in (\ref{eq:sys2global}) involves the aggregation of all the local weights which is challenging to obtain due to the cross-term caused by the low-rank model transmission, channel fading, and noise in the wireless communication systems as we will illustrate in Section \ref{sec:OTA}.
\end{enumerate}
In the following two subsections, we will present the solutions to overcome the above two challenges.
\subsection{Riemannian Gradient Descent to Update the Local Model Weights {in Step 1}}\label{sec:local}

Since there is no closed-form solution to the optimization in (\ref{eq:sys2local}), we can only seek the inexact solution. Here we propose the Riemannian gradient descent (RGD)  to obtain a local solution  based on the results from the last iteration. {      The procedures of RGD   contain the following two main steps:
\begin{itemize}
\item {\bf Step 1-1 (Riemannian Gradient Calculation):} Calculate the  Riemannian  gradient $\bm{g}^{(t)}_{k}$ for  Problem (\ref{eq:sys2local}) on the tangent space $\mathcal{T}_{\bm{\Theta}_k^{(t)}}\mathcal{M}_R$ of  $\bm{\Theta}_k^{(t)}$ on $\mathcal{M}_R$,
    \item {\bf Step 1-2 (Local Model Weights  Update):}  Move along the geodesic starting at $\bm{\Theta}_k^{(t)}$
 with direction $-\bm{g}^{(t)}_{k}$ and step size $\eta$ on the tangent space, and use retraction mapping $\mathcal{R}_{\bm{\Theta}_k^{(t)}}(-\eta^{(t)}\bm{g}^{(t)}_k)$ to update the local model weights $\bm{\Theta}_k^{(t+1)}$.
\end{itemize}
Fig (\ref{fig:RGD}) shows the geometric interpretation of the above two steps for RGD.
	\begin{figure}[H]
	\centering
	\includegraphics[width=0.8\linewidth]{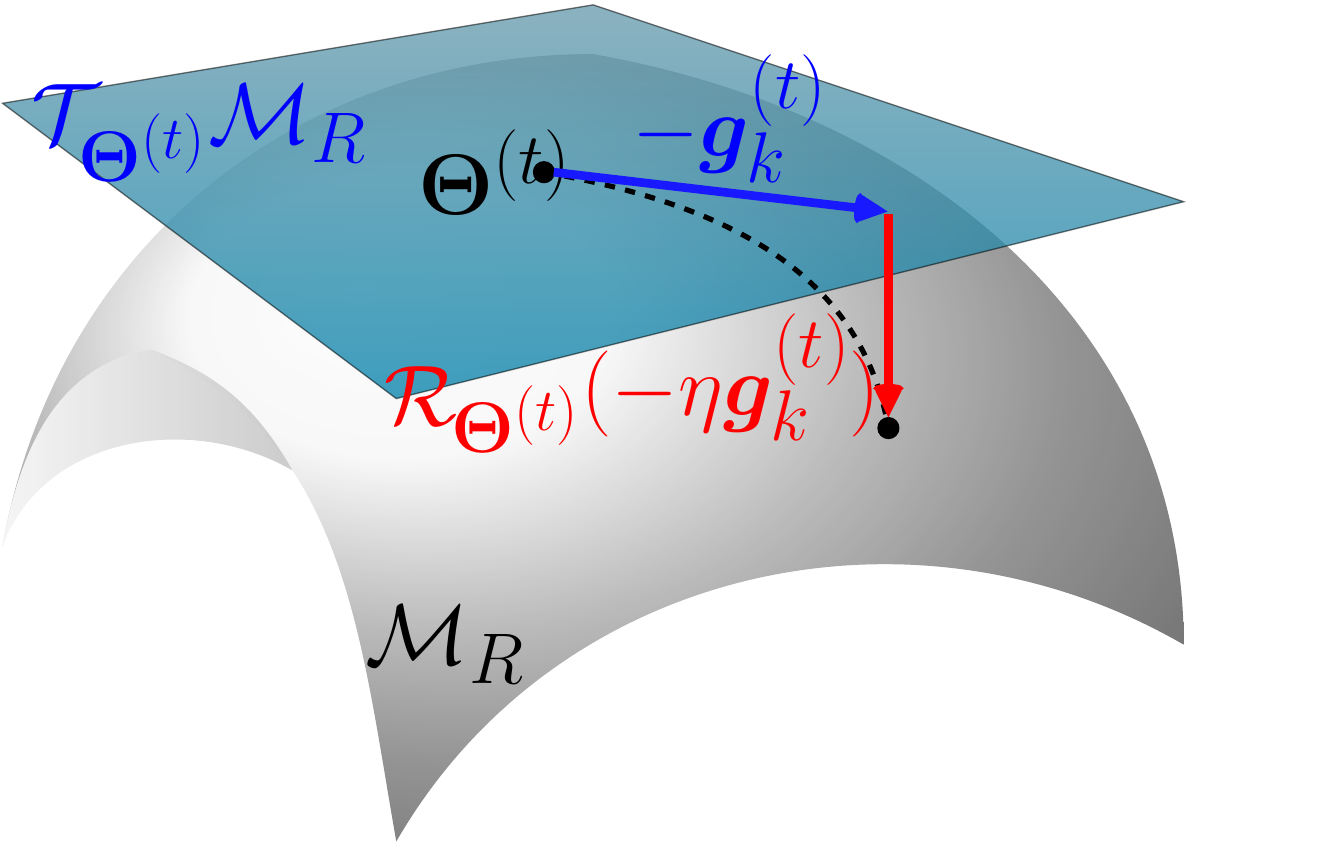}
	\caption{{      Illustration of RGD for solving Problem (\ref{eq:sys2local}).}}
	\label{fig:RGD}
\end{figure}}

\subsubsection{Riemannian Gradient Calculation}
To calculate the   Riemannian  gradient, $\bm{g}^{(t)}_{k}$,  at point $\bm{\Theta}_k^{(t)}$ for  Problem (\ref{eq:sys2local}), we first need the tangent space of $\mathcal{M}_R$ at $\bm{\Theta}_k^{(t)}$ as defined in \cite{wei2016guarantees}, {     which is expressed by the following Proposition \ref{thm:deftang}.
\begin{prop}[Tangent Space of $\mathcal{M}_R$ \cite{wei2016guarantees}]\label{thm:deftang}
Let $\bm{\Theta}_k^{(t)}$ be an arbitrary point on  $\mathcal{M}_R$ with compact  singular value decomposition (SVD) as
\begin{equation}
    SVD_{comp}(\bm{\Theta}_k^{(t)})=[\bm{U}^{(t)}_k,\bm{\Sigma}^{(t)}_k,\bm{V}^{(t)}_k],
\end{equation} where $\bm{U}^{(t)}_k \in \mathcal{ST}_{M}^{R}$,  $\bm{U}^{(t)}_k \in \mathcal{ST}_{N}^{R}$, and $\bm{\Sigma}^{(t)}_k$ is the diagonal matrix containing the singular values. Then, the  tangent space $\mathcal{T}_{\bm{\Theta}_k^{(t)}}\mathcal{M}_R$ of $\mathcal{M}_R$ at $\bm{\Theta}_k^{(t)}$ is given by 
\begin{equation}
\begin{aligned}
 \mathcal{T}_{\bm{\Theta}_k^{(t)}}\mathcal{M}_R&=\\
 &\Big\{\bm{U}^{(t)}_k(\bm{\Omega}_1)^T+\bm{\Omega}_2(\bm{V}^{(t)}_k)^T\\
 &\text{with } \forall \bm{\Omega}^{(t)}_1\in\mathbb{R}^{N\times R},\forall \bm{\Omega}^{(t)}_2\in\mathbb{R}^{M\times R}\Big\},
\end{aligned}
\end{equation}
\end{prop}
As derived  in Appendix \ref{der:rg}, the  Riemannian  gradient at $\bm{\Theta}_k^{(t)}\in \mathcal{M}_R $ for  Problem (\ref{eq:sys2local}) is given by

{\small\begin{equation}\label{eq:gradp1}
    \begin{aligned}
    \bm{g}^{(t)}_{k}=&\bm{U}^{(t)}_k(\bm{U}^{(t)}_k)^T\big(\nabla_{\bm{\Theta}_k} \frac{1}{KD}\sum_{l=1}^Df(\bm{\Theta}^{(t)}_k,\bm{\zeta}_{k,l})+\frac{\mu}{K}\bm{\Theta}^{(t)}_k-\frac{\mu}{K}\bm{\Theta}^{(t)}_0\big)\\
 &+\big(\nabla_{\bm{\Theta}_k} \frac{1}{KD}\sum_{l=1}^Df(\bm{\Theta}^{(t)}_k,\bm{\zeta}_{k,l})+\frac{\mu}{K}\bm{\Theta}^{(t)}_k-\frac{\mu}{K}\bm{\Theta}^{(t)}_0\big)\\
 &\times\bm{V}^{(t)}_k(\bm{V}^{(t)}_k)^T
 -\bm{U}^{(t)}_k(\bm{U}^{(t)}_k)^T\big(\nabla_{\bm{\Theta}_k} \frac{1}{KD}\sum_{l=1}^Df(\bm{\Theta}^{(t)}_k,\bm{\zeta}_{k,l})\\
 &+\frac{\mu}{K}\bm{\Theta}^{(t)}_k-\frac{\mu}{K}\bm{\Theta}^{(t)}_0\big)\bm{V}^{(t)}_k(\bm{V}^{(t)}_k)^T.\\
    \end{aligned}
\end{equation}}

Note that $\nabla \frac{1}{D}\sum_{l=1}^Df(\bm{\Theta}^{(t)}_k,\bm{\zeta}_{k,l})$ in (\ref{eq:gradp1}) can be approximated by the mini-batch version $\nabla \frac{1}{|\mathcal{D}^{(t)}_k|}\sum_{\bm{\zeta}_{k,l}\in\mathcal{D}^{(t)}_k}f(\bm{\Theta}^{(t)}_k,\bm{\zeta}_{k,l})$ to achieve a lower computation complexity,	where $\mathcal{D}^{(t)}_k\subseteq\mathcal{D}_k$ denotes the uniformly random selected mini-batch. Then, we can obtain the mini-batch Riemannian  gradient as follows,
{\small \begin{equation}\label{eq:gradappp1}
\begin{aligned}
&\bm{\hat g}^{(t)}_{k}=\\&\bm{U}^{(t)}_k(\bm{U}^{(t)}_k)^T\Big(\nabla_{\bm{\Theta}_k} \frac{1}{K|\mathcal{D}^{(t)}_k|}\sum_{\bm{\zeta}_{k,l}\in\mathcal{D}^{(t)}_k}f(\bm{\Theta}^{(t)}_k,\bm{\zeta}_{k,l})+\frac{\mu}{K}\bm{\Theta}^{(t)}_k\\
&-\frac{\mu}{K}\bm{\Theta}^{(t)}_0\Big)
 +\Big(\nabla_{\bm{\Theta}_k} \frac{1}{K|\mathcal{D}^{(t)}_k|}\sum_{\bm{\zeta}_{k,l}\in\mathcal{D}^{(t)}_k}f(\bm{\Theta}^{(t)}_k,\bm{\zeta}_{k,l})\\
 &\quad+\frac{\mu}{K}\bm{\Theta}^{(t)}_k-\frac{\mu}{K}\bm{\Theta}^{(t)}_0\Big)\bm{V}^{(t)}_k(\bm{V}^{(t)}_k)^T\\
 &-\bm{U}^{(t)}_k(\bm{U}^{(t)}_k)^T\Big(\nabla_{\bm{\Theta}_k} \frac{1}{K|\mathcal{D}^{(t)}_k|}\sum_{\bm{\zeta}_{k,l}\in\mathcal{D}^{(t)}_k}f(\bm{\Theta}^{(t)}_k,\bm{\zeta}_{k,l})\\
 &\quad+\frac{\mu}{K}\bm{\Theta}^{(t)}_k-\frac{\mu}{K}\bm{\Theta}^{(t)}_0\Big)\bm{V}^{(t)}_k(\bm{V}^{(t)}_k)^T.
 \end{aligned}
\end{equation}}}

\subsubsection{Local Model Weights  Update }
After obtaining the mini-batch Riemannian gradient, the device needs to move along a geodesic\footnote{A geodesic is commonly a curve representing  the shortest path  between two points on a Riemannian manifold.  It is a generalization of the notion of a "straight line" to a more general setting.} in the  negative Riemannian gradient direction over $\mathcal{M}_R$ to update the local model weights. However, moving along a geodesic requires the exponential map over $\mathcal{M}_R$, which is computationally expensive since it involves solving an
ordinary differential equation for which we do not have an analytical solution in general. Therefore, we adopt its {      second-order} approximation, the {\em retraction} operation \cite{absil2012projection} $\mathcal{R}_{\bm{\Theta}_k^{(t)}}(\cdot)$, to update the variable. Specifically, we obtain the updated model weights by

\begin{equation}\label{eq:update}
    \begin{aligned}
    \bm{\Theta}_k^{(t+1)}=&\mathcal{R}_{\bm{\Theta}_k^{(t)}}(-\eta^{(t)}\bm{\hat g}^{(t)}_k)\\
    =&\bm{U}_k^{(t+1)}\bm{\Sigma}^{(t+1)}_k(\bm{V}_k^{(t+1)})^T,
    \end{aligned}
\end{equation}
where
\begin{equation}\label{eq:svdr}
    \begin{aligned}
 [\bm{U}_k^{(t+1)},\bm{\Sigma}^{(t+1)}_k, (\bm{V}_k^{(t+1)})]=SVD_R(\bm{\Theta}^{(t)}-\eta^{(t)}\bm{\hat g}^{(t)}_k),\\
    \end{aligned}
\end{equation}
with $SVD_R(\cdot)$  the truncated SVD operation that only calculates the top-$R$ singular values and the corresponding singular vectors. Specifically, $\bm{\Sigma}^{(t+1)}_k$ is $R\times R$ diagonal with the $R$ largest singular values in its diagonal,  matrices $\bm{U}_k^{(t+1)}\in \mathbb{R}^{M\times R}$ and $\bm{V}_k^{(t+1)}\in \mathbb{R}^{N\times R}$ contain  the corresponding left and right singular vectors, respectively. {      The proximal method for solving the nuclear-norm-relaxation problem (\ref{prob:nuclear}) also involves an SVD, but it truncated the eigenvectors by a soft-thresholding via a complex relationship between the desired rank $R$ and the regularization parameter $\lambda$. To satisfy the required rank constraint, the correct hyperparameter may vary across different datasets or even different iterations, making it more challenging to implement in practice.}

\subsection{Factorized Local Model Transmission with OTA Aggregation to Update the Global Model Weights {in Step 2}}\label{sec:OTA}

After obtaining $ \bm{\Theta}_k^{(t+1)}$ via (\ref{eq:gradappp1}) and (\ref{eq:update}), we are able to update the global model weight matrix by
{\small\begin{equation}\label{eq:severupdate}
    \begin{aligned}
    (\bm{\Theta}_0^*)^{(t+1)}&=\underset{\bm{\Theta}_{0}\in\mathcal{M}_R}{\arg\min}\quad  \frac{\mu}{2K}\sum_{k=1}^K\|\bm{\Theta}_{0}-\bm{\Theta}^{(t+1)}_{k}\|_F^2\\
    &=\mathcal{P}_{\mathcal{M}_R}(\frac{1}{K}\sum_{k=1}^K\bm{\Theta}^{(t+1)}_{k}),
    \end{aligned}
\end{equation}}which requires  the server to aggregate the local models as $\frac{1}{K}\sum_{k=1}^K\bm{\Theta}^{(t+1)}_{k}$ and to broadcast the aggregated model to all  devices after a projection $\mathcal{P}_{\mathcal{M}_R}(\frac{1}{K}\sum_{k=1}^K\bm{\Theta}^{(t+1)}_{k})$. However, naively transmitting the original  local models $\bm{\Theta}^{(t+1)}_{k},\forall k\in[K]$  will bring a huge communication cost. Therefore, we propose to reduce the communication cost with a factorized local model transmission to leverage the low-rank property and with the OTA aggregation to reduce the communication cost. Detailed illustrations will be given in the remainder of this subsection.
  \subsubsection{Factorized Local Model Transmission at the device}\label{sec:fac}
According to the local model update in  (\ref{eq:update}), we can represent the local model weight matrix at each device $k$ by 
{\small\begin{equation}\label{eq:comp}
\begin{aligned}
\bm{\Theta}_k^{(t+1)}&=\tilde{\bm{U}}_k^{(t+1)}\big(\tilde{\bm{V}}_k^{(t+1)}\big)^T,\\
\text{with  }&\tilde{\bm{U}}_k^{(t+1)}=\bm{U}_k^{(t+1)}\sqrt{\bm{\Sigma}^{(t+1)}_k}\in \mathbb{R}^{M\times R},\\
&\tilde{\bm{V}}_k^{(t+1)}=\bm{V}_k^{(t+1)}\sqrt{\bm{\Sigma}^{(t+1)}_k}\in \mathbb{R}^{N\times R}.\\
\end{aligned}
\end{equation}}
Therefore, each device $k$ can transmit $\tilde{\bm{U}}_k^{(t+1)}$ and $\tilde{\bm{V}}_k^{(t+1)}$ instead of $ \bm{\Theta}_k^{(t+1)}$. As such, the  communication cost for each device is reduced from $\mathcal{O}(MN)$  to $\mathcal{O}((M+N)R)$ in each aggregation.

However, the factorization in (\ref{eq:comp}) will lead to the {\em cross-term issue} when being integrated with OTA aggregation 
as explained by the following expression,
{\small\begin{equation}\label{eq:OTA0}
    \begin{aligned}
    &\underbrace{\sum_{k=1}^K\tilde{\bm{U}}_{k}^{(t+1)}}_{\text{OTA Aggregated } \tilde{\bm{U}}_k^{(t+1)}}\times \underbrace{\sum_{k=1}^K\big(\tilde{\bm{V}}_{k}^{(t+1)}\big)^T}_{\text{OTA Aggregated }  \big(\tilde{\bm{V}}_k^{(t+1)}\big)^T}
     \end{aligned}
\end{equation}}
{\small\begin{equation}\label{eq:OTA1}
    \begin{aligned}
   =&\underbrace{\sum_{k=1}^K\tilde{\bm{U}}_{k}^{(t+1)}\big(\tilde{\bm{V}}_{k}^{(t+1)}\big)^T}_{\text{Desired Aggregated Model}}+ \underbrace{\sum_{k=1, j\neq k}^K\tilde{\bm{U}}_{k}^{(t+1)}\big(\tilde{\bm{V}}_{j}^{(t+1)}\big)^T}_{\text{Cross Term}}
      \end{aligned}
\end{equation}}
{\small\begin{equation}\label{eq:OTA}
    \begin{aligned}
    =&\underbrace{\sum_{k=1}^K\bm{\Theta}^{(t+1)}_{k}}_{\text{Desired Aggregated Model}}+ \underbrace{\sum_{k=1, j\neq k}^K\tilde{\bm{U}}_{k}^{(t+1)}\big(\tilde{\bm{V}}_{j}^{(t+1)}\big)^T}_{\text{Cross Term}}. 
    \end{aligned}
\end{equation}}(\ref{eq:OTA}) shows that if the local devices directly transmit the two factorized  small matrices to the server and the server naively aggregates them over the air, there will be an undesired cross-term signal.

Fortunately, the unbiased estimation of the desired accurate aggregated model $\sum_{k=1}^K\bm{\Theta}^{(t+1)}_{k}$ is sufficient for the convergence of the algorithm in the stochastic setting, as we will show in Section \ref{sec:conv}. Therefore, we propose the  RLC-based precoder at each device for the server to obtain an unbiased estimation of the accurate aggregated model. Specifically, we define the RLC-based precoder for the $k$-th  device as follows.

\begin{defn}\label{thm:defRLC}(The RLC-based Precoder)
A RLC-based Precoder for the $k$-th device, denoted by $\bm{F}_k\in\mathbb{R}^{R\times R}$, is constructed by drawing each entry, ${F}_k[i,j]$,  independently from a  Radamacher  distribution  given by
\begin{equation}
    P\Big({F}_k[i,j]=\frac{1}{\sqrt{R}}\Big) =P\Big({F}_k[i,j]=-\frac{1}{\sqrt{R}}\Big) =\frac{1}{2}.
\end{equation}
Note that the precoders among all devices  are also generated in an i.i.d. manner.
\end{defn}
Adopting the RLC-based precoder, the final transmitted signals at the $t$-th training iteration from  the $k$-th  device
are
\begin{equation}\label{eq:LRCP}
\begin{aligned}
\bm{S}^{(t+1)}_{Uk}=\tilde{\bm{U}}_k^{(t+1)}\bm{F}_k,\quad
\bm{S}^{(t+1)}_{Vk}=\big(\tilde{\bm{V}}_k^{(t+1)}\bm{F}_k\big)^T,
\end{aligned}
\end{equation}
where $\tilde{\bm{U}}_k^{(t+1)}$ and $\tilde{\bm{V}}_k^{(t+1)}$ are obtained by (\ref{eq:comp}), $\bm{F}_k$ is from Definition \ref{thm:defRLC}. It is easy to verify that after adopting the RLC-based precoders, the expected value of the cross-term is zero.

\subsubsection{OTA Aggregation at the Server}\label{sec:otasever}

After each device, $k$  transmits its local signals with transmission power
coefficient $p_{k}^{(t+1)}$, the server will receive the raw signals
 {\small \begin{equation}\label{eq:SEVERRX}
 \begin{aligned}
&\bm{Y}_U^{(t+1)}=\sum^K_{k=1}h_{k}^{(t+1)}p_{k}^{(t+1)}\bm{S}_{Uk}^{(t+1)} + \bm{Z}_U^{(t+1)},\\
&\bm{Y}_V^{(t+1)}=\sum^K_{k=1}h_{k}^{(t+1)}p_{k}^{(t+1)}\bm{S}_{Vk}^{(t+1)} + \bm{Z}_V^{(t+1)}.\\
 \end{aligned}
 \end{equation}}
 To obtain the aggregation of the transmitted  signals, $\bm{X}_U^{(t+1)}=\sum_{k=1}^K\bm{S}_{Uk}^{(t+1)}$ and $\bm{X}_V^{(t+1)}=\sum_{k=1}^K\bm{S}_{Vk}^{(t+1)}$ from the received signal $\bm{Y}^{(t+1)}_U$ and  $\bm{Y}^{(t+1)}_V$, the server adopts  the LS  estimation method  to obtain the estimates $\hat{\bm{X}_U}^{(t+1)}$ and $\hat{\bm{X}_V}^{(t+1)}$ by
{\small\begin{equation}\label{eq:LS}
\begin{aligned}
\hat{\bm{X}}_U^{(t+1)}
	=&\frac{Re(\bm{Y}_U^{(t+1)})}{\rho^{(t+1)}}
	=&\frac{1}{\rho^{(t+1)}}\sum_{k=1}^K \bm{S}^{(t+1)}_{Uk}Re(p_{k}^{(t+1)}h_{k}^{(t+1)})\\
	&&+\frac{Re(\bm{Z}_U^{(t+1)})}{\rho^{(t+1)}},\\
	\hat{\bm{X}}_V^{(t+1)}
	=&\frac{Re(\bm{Y}_V^{(t+1)})}{\rho^{(t+1)}}
	=&\frac{1}{\rho^{(t+1)}}\sum_{k=1}^K \bm{S}^{(t+1)}_{Vk}Re(p_{k}^{(t+1)}h_{k}^{(t+1)})\\
	&&+\frac{Re(\bm{Z}_V^{(t+1)})}{\rho^{(t+1)}}.
	\end{aligned}
	\end{equation}}where $\rho^{(t+1)}$ is the receiving power coefficient according to the specific unbiased power policy.
	\begin{remark}[Unbiased Power {Control} Policy]
	{The proposed scheme in (\ref{eq:SEVERRX}) and (\ref{eq:LS}) can be used with} any unbiased power {control} policy. An unbiased power {control} policy refers to  the power coefficients $\{p_k^{(t+1)}\}_{k=1}^K$ and $\rho^{(t+1)}$, which can guarantee  $\mathbb{E}_{\{\bm{F}_{k}\}_{k=1}^K,\{h_{k}^{(t+1)}\}_{k=1}^K,\bm{Z}_U^{(t+1)},\bm{Z}_V^{(t+1)}}[\hat{\bm{X}}_U^{(t+1)}\times\hat{\bm{X}}_V^{(t+1)}]=\mathbb{E}_{\{\bm{F}_{k}\}_{k=1}^K}[{\bm{X}}_U^{(t+1)}\times {\bm{X}}_V^{(t+1)}]$. {      Note that the power coefficients $p_k^{(t+1)}$ usually rely on  CSI.} Many widely used power {control} policies, such as GBMA \cite{sery2020analog}, CI \cite{xing2021federated}, are unbiased.   For example, with the GBMA power policy, we have
	\begin{equation}\label{eq:GBMA}
	    \begin{aligned}
	    p_k^{(t+1)}&=\sqrt{\gamma}e^{j-\angle{h^{(t+1)}_k}},\\
	    \rho^{(t+1)}&={\sqrt{\gamma}}.
	    \end{aligned}
	\end{equation}\
		{  Therefore, we can obtain
			\begin{equation}\label{eq:GBMAunbia}\small
	    \begin{aligned}
	  & \mathbb{E}_{\{\bm{F}_{k}\}_{k=1}^K,\{h_{k}^{(t+1)}\}_{k=1}^K,\bm{Z}_U^{(t+1)},\bm{Z}_V^{(t+1)}}[\hat{\bm{X}}_U^{(t+1)}\times\hat{\bm{X}}_V^{(t+1)}]\\
	   =&\mathbb{E}_{\{\bm{F}_{k}\}_{k=1}^K,\{h_{k}^{(t+1)}\}_{k=1}^K,\bm{Z}_U^{(t+1)},\bm{Z}_V^{(t+1)}}\bigg[\Big(\sum_{k=1}^K \bm{S}^{(t+1)}_{Uk}|h_{k}^{(t+1)}|\\
	   &\quad+\frac{Re(\bm{Z}_U^{(t+1)})}{\sqrt{\gamma}}\Big)
	   \times \Big(\sum_{k=1}^K \bm{S}^{(t+1)}_{Vk}|h_{k}^{(t+1)}|+\frac{Re(\bm{Z}_V^{(t+1)})}{\sqrt{\gamma}}\Big)\bigg]\\
	     =&\mathbb{E}_{\{\bm{F}_{k}\}_{k=1}^K,\{h_{k}^{(t+1)}\}_{k=1}^K}\bigg[\Big(\sum_{k=1}^K \bm{S}^{(t+1)}_{Uk}|h_{k}^{(t+1)}|\Big)
	   \times \Big(\sum_{k=1}^K \bm{S}^{(t+1)}_{Vk}|h_{k}^{(t+1)}|\Big)\bigg]\\
	   =&\mathbb{E}_{\{\bm{F}_{k}\}_{k=1}^K}\bigg[\Big(\sum_{k=1}^K \bm{S}^{(t+1)}_{Uk}\Big)
	   \times \Big(\sum_{k=1}^K \bm{S}^{(t+1)}_{Vk}\Big)\bigg]\\
	   =&\mathbb{E}_{\{\bm{F}_{k}\}_{k=1}^K}[{\bm{X}}_U^{(t+1)}\times {\bm{X}}_V^{(t+1)}].
	    \end{aligned}
	\end{equation}}
	\end{remark}

   After obtaining  $\hat{\bm{X}}_U^{(t+1)}$ and $\hat{\bm{X}}_V^{(t+1)}$ from (\ref{eq:LS}), the server will obtain the aggregated model by
    \begin{equation}\label{eq:aggest}
\begin{aligned}
{\bm{\Theta}}^{(t+1)}_{0}=&\mathcal{P}_{\mathcal{M}_R}(\frac{1}{\sqrt{K}}\hat{\bm{X}}_U^{(t+1)}\times\frac{1}{\sqrt{K}}\hat{\bm{X}}_V^{(t+1)}).\\
	\end{aligned}
	\end{equation}
{In an error-free communication system, one can directly obtain the optimal aggregated model $({\bm{\Theta}}_{0}^*)^{(t+1)}$ via (\ref{eq:severupdate}). Considering the random noise in real communication systems, we propose to use ${\bm{\Theta}}^{(t+1)}_{0}$ in (\ref{eq:aggest})  as an approximation of  ${\bm{\Theta}}^{(t+1)}_{0}$. We will show in Section \ref{sec:conv} that ${\bm{\Theta}}^{(t+1)}_{0}$ is sufficient to guarantee   the convergence of the overall training in expectation.}

Using the proposed factorized local model transmission with the OTA aggregation, the communication cost can be reduced  from $\mathcal{O}(KMN)$ to $\mathcal{O}(R(M+N))$.

\subsection{Overall Implementation}
\subsubsection{The Proposed FedRLR Algorithm}
\begin{figure}[htbp]
	\centering
	\includegraphics[width=1.0\linewidth]{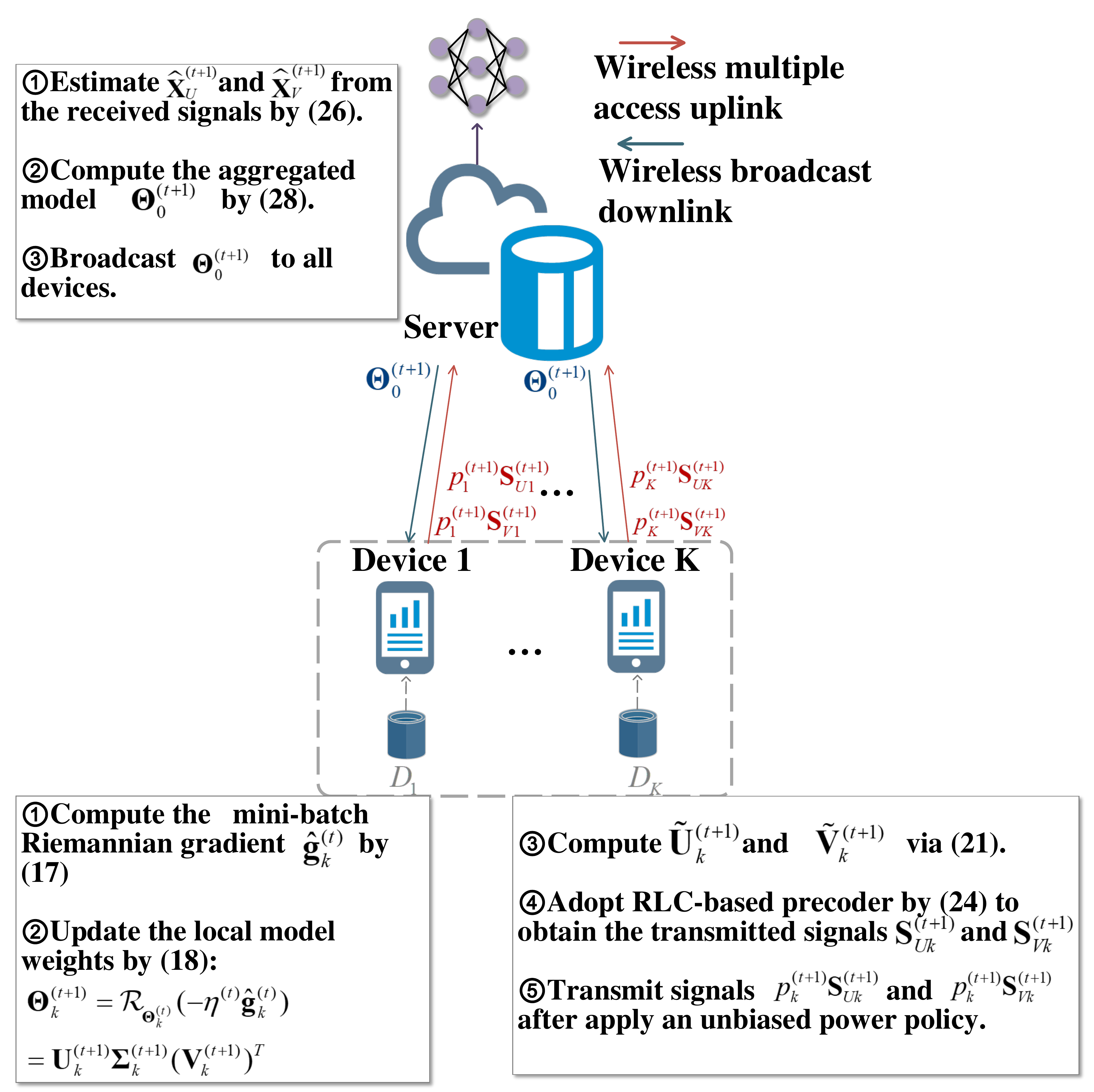}
	\caption{The implementation process of the proposed FedRLR algorithm. }
	\label{fig:RLAVG}
\end{figure}

Combining the results in Section \ref{sec:local} and  Section  \ref{sec:OTA}, we summarize the proposed federated Riemannian-based alternating low-rank model compression with OTA aggregation, termed FedRLR, in Algorithm \ref{alg:genFWODL} and Fig. \ref{fig:RLAVG}.
\begin{algorithm}[htbp]\label{alg:genFWODL}
	\caption{Federated Riemannian-based Alternating Optimization  with the OTA Aggregation for Low-Rank Model Compression (FedRLR)}
	\KwData{Input data $\bm{\zeta}_{l_k},\forall k\in[K], l_k\in [D]$, required rank $R$, total number of training rounds $T_{iter}$.} 
	\KwResult{$\bm{\Theta}^{(T_{iter})}_{k}\in \mathcal{M}_R,\forall k\in[K]$ }
	Initialization: {  $\bm{\Theta}^{(0)}_{k}=\bm{\Theta}_{0}^{(0)},[\bm{U}_k^{(0)},\bm{\Sigma}_k^{(0)},(\bm{V}_k^{(0)})^T]=SVD_{R}(\bm{\Theta}^{(0)}_{k}),\forall k\in[K]$ where $\bm{\Theta}_{0}^{(0)}$ is a random point on  $\mathcal{M}_R$}\;	
	\For{$ t = 0,1,2,\ldots,T_{iter}$}
	{	
	 {\bf {\text Each device} $k$ :}
	 {\text {\em 1. Calculates the Riemannian mini-batch gradient}}\\{$\bm{\hat g}^{(t)}_{k}$ according to  (\ref{eq:gradappp1}) with $
		{\bm{\Theta}}^{(t)}_{0}$, $
		{\bm{\Theta}}^{(t)}_{k}$ and uniformly randomly sampled mini-batch local data. }\\
		{\text {\em 2. Updates the local model weights $\bm{\Theta}^{(t+1)}_{k}$}} \\\text{{\em according to (\ref{eq:update})}}\\
	{\text {\em 3. Prepares the compressed representation}}\\{\text {\em of the local model weights via (\ref{eq:comp}):}}\\
{$ {\bm{\tilde U}}_k^{(t+1)}=\bm{U}_k^{(t+1)}\sqrt{\bm{\Sigma}^{(t+1)}_k},$\\
${\bm{\tilde V}}_k^{(t+1)}=\bm{V}_k^{(t+1)}\sqrt{\bm{\Sigma}^{(t+1)}_k}$},
\\
with $\bm{\Theta}_k^{(t+1)}=\tilde{\bm{U}}_k^{(t+1)}\big(\tilde{\bm{V}}_k^{(t+1)}\big)^T$.\\
{\text {\em 4. Adopts RLC-based precoder}}\\   {\text {\em to obtain the transmitted signals by (\ref{eq:LRCP}):}}\\
		{$ \bm{S}^{(t+1)}_{Uk}=\tilde{\bm{U}}_k^{(t+1)}\bm{F}_k,$\\
        $\bm{S}^{(t+1)}_{Vk}=\big(\tilde{\bm{V}}_k^{(t+1)}\bm{F}_k\big)^T$}.\\
		{\text {\em 5. Transmits the signals using unbiased power policy:}}\\
		{ \text{Device} $k$ \text{transmits} $p_{k}^{(t+1)}\bm{S}_{Uk}^{(t+1)}$ and $p_{k}^{(t+1)}\bm{S}_{Vk}^{(t+1)}$ to the server}.

		{\bf \text At  the server:}\\
		{\text{\em 1.  Receives signal via (\ref{eq:SEVERRX})}}
			{\text{\em 2. Calculates the LS estimation  from the received }}\\\text{{\em signal by (\ref{eq:LS}). }}\\
		{\text{\em 3.  Calculates the aggregated model ${\bm{\Theta}}^{(t+1)}_{0}$}}
			 \\	{\text{\em according to (\ref{eq:aggest}). }}\\
		{\text{\em 4.  Broadcasts the aggregated model $
		{\bm{\Theta}}^{(t+1)}_{0}$
			to all devices. }}\\
			}
\end{algorithm}

\subsubsection{Computational Complexity Analysis}
{      We analyze the
computational complexity in terms of the number of the floating-point operations (FLOPs) \cite{hunger2005floating}. The main operations of the proposed algorithm  includes matrix multiplication,
 and truncated SVD of a matrix. The number of FLOPs for the proposed algorithm  is summarized in
Table \ref{tab:comple}.
\begin{table}[ht]
	\caption{Per-iteration Complexity Analysis}
	\label{tab:comple}
	\centering
	\begin{tabular}{lll}
		\bottomrule
			\rowcolor{light-gray} {}&	{Operation} & {Number of FLOPs}\\
			\toprule
	     \multirow{4}{*}{Device} &Riemannian gradient&$\mathcal{O}(MNR+R^2N+R^2M)$\\
  &Retraction& $\mathcal{O}(MNR)$ \cite{alameddin2019toward,absil2015low}\\
  &Compressed representation&$\mathcal{O}(MR+NR)$\\
    &RLC precoding&$\mathcal{O}(MR^2+NR^2)$\\
     \hline
 \multirow{2}{*}{Server}& LS estimation&$\mathcal{O}(MR+NR)$\\
    &Aggregated model&$\mathcal{O}(MNR)$\\
		\bottomrule
	\end{tabular}
\end{table}
The overall computation complexity of our algorithm is at the order of $\mathcal{O}(MNR)$, which is in the same order as that of gradient descent  with a scaling factor $R$. Since in low-rank model compression, $R$ is much smaller than $M$ or $N$, we believe that the computational overhead of Riemannian gradient descent is manageable for mobile devices and the server, and will not result in heavy computation overheads.}
\section{Convergence Analysis}\label{sec:conv}
In this section, we will show that the proposed algorithm can converge to a KKT point of Problem $\mathscr{P}2$. For  notational simplicity, we  denote the stacked local model weights as $\vec{\bm{\Theta}}=(\bm{\Theta}_1^T,\bm{\Theta}_2^T,\ldots,\bm{\Theta}_K^T)^T$, and $\vec{\bm{\Theta}^*_0}=\bm{\Theta}^*_0\otimes \bm{1}_K$, where $\bm{\Theta}^*_0 \triangleq\mathcal{P}_{\mathcal{M}_R}(\frac{1}{K}\sum_{k=1}^K\bm{\Theta}_{k})$.

 {Since Problem $\mathscr{P}2$ is a problem over the smooth Riemannian manifold, the definition of the KKT points in Euclidean space is generalized to the smooth Riemannian manifold  for Problem $\mathscr{P}2$ according to \cite{bergmann2019intrinsic} and is given as follows.}
\begin{defn}({KKT Point of Problem $\mathscr{P}2$ over the Smooth Riemannian Manifold $ \mathcal{M}_R$})\label{def:KKT}\\
A KKT point of Problem $\mathscr{P}2$ is the point that satisfies the following conditions:
\begin{enumerate}
    \item (Primal Feasibility)
    {\small\begin{equation}
        \bm{\Theta}_k=\bm{\Theta}_0, \quad  \bm{\Theta}_k \in \mathcal{M}_R, \forall k\in[K], \bm{\Theta}_0 \in \mathcal{M}_R. 
    \end{equation}}
    
    \item (Stationarity of the Lagrangian) 
    {\small\begin{equation}
        \begin{aligned}
        &grad_{\vec{\bm{\Theta}}}L_{\mathscr{P}2}(\vec{\bm{\Theta}},\bm{\Theta}_0,\vec{\bm{\Upsilon}})\\
       =&  grad_{\vec{\bm{\Theta}}}\bigg(\frac{1}{K}\sum_{k=1}^K\frac{1}{D}\sum_{l=1}^Df(\bm{\Theta}_k,\bm{\zeta}_{k,l})+\sum_{k=1}^K\langle \bm{\Upsilon}_k,\bm{\Theta}_k-\bm{\Theta}_0\rangle\bigg)\\
       =&\bm{0},
        \end{aligned}\label{eq:lag1}
    \end{equation}}
      {\small\begin{equation}
        \begin{aligned}
        &grad_{{\bm{\Theta}_0}}L_{\mathscr{P}2}(\vec{\bm{\Theta}},\bm{\Theta}_0,\vec{\bm{\Upsilon}})\\
       =&  grad_{{\bm{\Theta}_0}}\bigg(\frac{1}{K}\sum_{k=1}^K\frac{1}{D}\sum_{l=1}^Df(\bm{\Theta}_k,\bm{\zeta}_{k,l})+\sum_{k=1}^K\langle \bm{\Upsilon}_k,\bm{\Theta}_k-\bm{\Theta}_0\rangle\bigg)\\
       =&\bm{0},
        \end{aligned}\label{eq:lag2}
    \end{equation}}
    where $\vec{\bm{\Upsilon}}=(\bm{\Upsilon}_1^T,\bm{\Upsilon}_2^T,\ldots,\bm{\Upsilon}_K^T)^T$ is the  Lagrangian multiplier.
\end{enumerate}
\end{defn}
\begin{remark}
In general constraint problems, there will be two other conditions, namely, complementary slackness and dual feasibility. However,  since there is no inequality constraint in $\mathscr{P}2$, the {complementary slackness} and  {dual feasibility} will automatically hold. Therefore, we omit these two conditions in Definition \ref{def:KKT} to improve the readability.
\end{remark}
Then, we present the analysis by   introducing the following assumptions.
\begin{assum}\label{thm:ass}
\quad
\begin{enumerate}

    \item (Bounded Local Mini-batch Euclidean Gradient)\\
    The uniform upper bound of $\|\nabla_{\bm{\Theta}_k} \frac{1}{|\mathcal{D}^{(t)}_k|}\sum_{\bm{\zeta}_{k,l}\in\mathcal{D}^{(t)}_k}f(\bm{\Theta}^{(t)}_k,\bm{\zeta}_{k,l})\|_F$ is $B_1$, i.e., for all $k$ and $t$, we have
    {\small\begin{equation}
    \begin{aligned}
   \|\nabla_{\bm{\Theta}_k} \frac{1}{|\mathcal{D}^{(t)}_k|}\sum_{\bm{\zeta}_{k,l}\in\mathcal{D}^{(t)}_k}f(\bm{\Theta}^{(t)}_k,\bm{\zeta}_{k,l})\|_F\leq B_1.
    \end{aligned}
    \end{equation}}
  
    \item (Lipschitz Local Task-driven Loss Function)\\
    The local task-driven loss $f(\bm{\Theta}_k,\bm{\zeta}_{k,l})$ has $L$-Lipschitz continuous Euclidean gradient, i.e., for all $k$ and $t$, we have
      {\small \begin{equation}
    \begin{aligned}
    &\Bigg|\frac{1}{D}\sum_{l=1}^Df(\bm{\Theta}_k,\bm{\zeta}_{k,l})-\bigg[\frac{1}{D}\sum_{l=1}^Df(\bm{\Theta}^{'}_k,\bm{\zeta}_{k,l})\\
    &+\langle\nabla_{\bm{\Theta}_k} \frac{1}{D}\sum_{l=1}^Df(\bm{\Theta}^{'}_k,\bm{\zeta}_{k,l}),\bm{\Theta}_k-\bm{\Theta}^{'}_k\rangle\bigg]\Bigg|\\
    &\leq \frac{L}{2}\|\bm{\Theta}_k-\bm{\Theta}^{'}_k\|_F^2,
    \end{aligned}
    \end{equation}}
    $\forall \bm{\Theta}_k, \bm{\Theta}^{'}_k\in\mathcal{M}_R$.
    \item (Unbiased Local Mini-batch Euclidean Gradient)\\
    The local mini-batch  Euclidean gradient is unbiased, i.e., for all $k$ and all $t$, we have
          {\small \begin{equation}
    \begin{aligned}
   &\mathbb{E}\Bigg[\nabla_{\bm{\Theta}_k}\frac{1}{|\mathcal{D}^{(t)}_k|}\sum_{\bm{\zeta}_{k,l}\in\mathcal{D}^{(t)}_k}f(\bm{\Theta}^{(t)}_k,\bm{\zeta}_{k,l})\Bigg]\\
   &=\nabla_{\bm{\Theta}_k} \frac{1}{D}\sum_{l=1}^Df(\bm{\Theta}^{(t)}_k,\bm{\zeta}_{k,l}).
    \end{aligned}
    \end{equation}}

    \item (Bounded Variance of the Local Stochastic Euclidean  Gradient)\\
    The local stochastic Euclidean gradient has bounded variance, i.e., for all $k$ and all $t$, we have
         {\small    \begin{equation}
    \begin{aligned}
   &\mathbb{E}\Bigg[\bigg\|\nabla_{\bm{\Theta}_k}f(\bm{\Theta}^{(t)}_k,\bm{\zeta}_{k,l})\\
   &\quad-\nabla_{\bm{\Theta}_k} \frac{1}{D}\sum_{l=1}^Df(\bm{\Theta}^{(t)}_k,\bm{\zeta}_{k,l})\bigg\|_F^2\Bigg]
   \leq\sigma_g^2.
    \end{aligned}
    \end{equation}}
\end{enumerate}
\end{assum}

{     
\begin{remark}
Assumption \ref{thm:ass} is widely used in  convergence analysis in machine learning or federated learning literature. Many commonly-used local task-driven loss functions $f(\bm{\Theta}^{(t)}_k,\bm{\zeta}_{k,l})$ satisfy the above Assumption \ref{thm:ass}, such as the cross-entropy loss for classification \cite{zhang2018generalized}, mean squared error loss for regression \cite{hastie2009elements}, and the KL divergence loss for distributional approximation \cite{kim2021comparing}, etc. Specifically, the bounded assumptions (assumptions 1) and 4)) are usually satisfied due to the bounded training data.  Assumption 2) means that the local loss function does not change too rapidly. Assumption 3) can be satisfied by sampling the training data in an unbiased way.

\end{remark}}
To show that the proposed algorithm can converge to a KKT point of  $\mathscr{P}2$, we will show that the limiting points of the proposed algorithm satisfy the primal feasibility and  the stationarity of the Lagrangian of  $\mathscr{P}2$. To start with, we present some technical lemmas that are required for the final convergence theory. 
First, we show the Lipschitz continuous inequality in the rank-$R$ manifold in Lemma \ref{lem:fixlip}.
\begin{lemma}(Lipschitz-type Inequality over $\mathcal{M}_R$)\label{lem:fixlip}
Under Assumption 1,  let $\bm{\Theta}_k$ and $\bm{\Theta}^{'}_k$ be two different points on $\mathcal{M}_R$, we have
{\small\begin{equation}
    \begin{aligned}
    &\Bigg|\frac{1}{D}\sum_{l=1}^Df(\bm{\Theta}_k,\bm{\zeta}_{k,l})-\bigg[\frac{1}{D}\sum_{l=1}^Df(\bm{\Theta}^{'}_k,\bm{\zeta}_{k,l})\\
    &+\langle grad_{\bm{\Theta}_k} \frac{1}{D}\sum_{l=1}^Df(\bm{\Theta}^{'}_k,\bm{\zeta}_{k,l}),\bm{\Theta}_k-\bm{\Theta}^{'}_k\rangle\bigg]\Bigg|\\
    &\leq \frac{L+6B_1}{2}\|\bm{\Theta}_k-\bm{\Theta}^{'}_k\|_F^2.
    \end{aligned}
    \end{equation}}
\end{lemma}
\begin{proof}
See Appendix \ref{proof:lem2}.
\end{proof}
With Lemma \ref{lem:fixlip}, we are able to show that the limiting points of the proposed algorithm satisfy the primal feasibility and  the stationarity of the Lagrangian of  $\mathscr{P}2$ by the following two lemmas.
\begin{lemma}(Converge to {a} Primal-Feasible Point {of $\mathscr{P}2$})\label{lem:convconsen}
Under Assumption \ref{thm:ass}, let the stepsize $\eta^{(t)}>0$ in (\ref{eq:update}) satisfy 
{\small\begin{equation}
    \begin{aligned}
       \sum_{t=0}^{\infty}\eta^{(t)}=\infty,\quad & \sum_{t=0}^{\infty}(\eta^{(t)})^2<\infty,\quad & \underset{t\to \infty}{\lim}\frac{\eta^{(t+1)}}{\eta^{(t)}}=1\\
    \end{aligned}
\end{equation}}
and the penalty parameter $\mu^{(t)}$ satisfy 
{\small\begin{equation}
    \mu^{(t)}=\frac{c_1}{\eta^{(t)}}
\end{equation} with $0<c_1<1$. For all $t$, we have 
\begin{equation}
    \begin{aligned}
      \mathbb{E}[\frac{ \|\vec{\bm{\Theta}}^{(t)}-\vec{\bm{\Theta}_0^*}^{(t)}\|_F^2}{K}]\leq c_2\frac{\sigma^2_z}{\gamma}B^2_1(\eta^{(t)})^2,
    \end{aligned}
\end{equation}}
with some absolute constant $c_2$.
\end{lemma}
\begin{proof}
See Appendix \ref{proof:lem3}.
\end{proof}
{     
\begin{remark}
    From Lemma \ref{lem:convconsen}, we can obtain that the  proposed FedRLR converges to a primal feasible point at a rate of $\mathcal{O}((\eta^{(t)})^2)$. Due to the relationship $\mu^{(t)}=\frac{c_1}{\eta^{(t)}}$, we can see that as the proportion of the penalty term gets larger, the local model weights gradually consensus.
\end{remark}}

\begin{lemma}(Converge to {a} Stationary Point of the Lagrangian in {(\ref{eq:lag1}) and (\ref{eq:lag2}))}\label{lem:stationary}
Under Assumption \ref{thm:ass}  and the conditions in Lemma \ref{lem:convconsen},  for any $\epsilon\geq\frac{2L+6B_1}{3}$, we have
{\small\begin{equation}\label{eq:convrate}
    \begin{aligned}
    &\inf_{1\leq t\leq T_{iter}}\mathbb{E}\Big[\|grad_{\vec{\bm{\Theta}}}L_{\mathscr{P}2}(\vec{\bm{\Theta}^{(t)}},\bm{\Theta}^{(t)}_0,\vec{\bm{\Upsilon}})\|_F^2\Big]\\
    \leq&\frac{ C^*+\sum_{t=0}^{T_{iter}}2(\eta^{(t)})^2\frac{\sigma^2_g}{D_m}
+2\frac{c^2_1c_2\sigma^2_zB^2_1(\eta^{(t)})^2}{K\gamma}}{\sum_{t=0}^{T_{iter}}\eta^{(t)}-(L+6B_1+3\epsilon)(\eta^{(t)})^2-\frac{1}{2\epsilon}},\\
    \end{aligned}
\end{equation}}
where $C^*=|\sup\frac{1}{K}\sum_{k=1}^K\frac{1}{D}\sum_{l=1}^Df(\bm{\Theta}_k,\bm{\zeta}_{k,l})-\inf \frac{1}{K}\sum_{k=1}^K\frac{1}{D}\sum_{l=1}^Df(\bm{\Theta}_k,\bm{\zeta}_{k,l})|$, $D_m=|\mathcal{D}^{(t)}_k|,\forall t, k$ and
{\small\begin{equation}\label{eq:convconst}
    \begin{aligned}
 &\mathbb{E}\Big[\|grad_{{\bm{\Theta}_0}}L_{\mathscr{P}2}(\vec{\bm{\Theta}},\bm{\Theta}_0,\vec{\bm{\Upsilon}})\|_F^2\Big]=0.
     \end{aligned}
\end{equation}}
\end{lemma}
\begin{proof}
See Appendix \ref{proof:lem4}.
\end{proof}
{      \begin{remark}
    From Lemma \ref{lem:stationary}, we can obtain that a larger minibatch size $D_m$ or a larger number of devices $K$ will lead to a faster convergence rate. 
\end{remark}}
Finally, combining Lemma \ref{lem:convconsen} and Lemma \ref{lem:stationary}, we can obtain the convergence theory as follows.

\begin{theorem}(Converge to {a KKT Point of $\mathscr{P}2$})\label{thm:converKKT}
Under Assumption \ref{thm:ass}, let the stepsize $\eta^{(t)}>0$ in (\ref{eq:update}) satisfy 
{\small\begin{equation}
    \begin{aligned}
       \sum_{t=0}^{\infty}\eta^{(t)}=\infty,\quad &  \sum_{t=0}^{\infty}(\eta^{(t)})^2<\infty,\quad & \underset{t\to \infty}{\lim}\frac{\eta^{(t+1)}}{\eta^{(t)}}=1\\
    \end{aligned}
\end{equation}}
and the penalty parameter $\mu^{(t)}$ satisfy 
\begin{equation}
    \mu^{(t)}=\frac{c_1}{\eta^{(t)}}
\end{equation} with $0<c_1<1$, {the proposed algorithm in Algorithm \ref{alg:genFWODL}} converges to the KKT point of Problem $\mathscr{P}2$ in expectation.
\end{theorem}

\begin{proof}
	Please refer to Appendix \ref{proof:thm1} for the proof.
\end{proof}

\section{Experiments}\label{sec:exp} 
In this section, we present simulation results and the
corresponding discussions for our proposed federated low-rank model compression scheme, FedRLR. In addition,  performance comparisons with various  baselines are also given. All the experiments are conducted in Python 3.7 with Pytorch 1.8.1.
\subsection{List of  Methods}\label{sec:list}
\begin{itemize}
	\item {\bf Benchmark}: The devices  collaboratively train a shared model without model compression under the help of a central server using the standard FedAvg \cite{mcmahan2017communication} in the traditional digital way.
	\item{\bf Proposed 1}: In this method, the local device training and the server aggregation adopt the proposed scheme in Algorithm (\ref{alg:genFWODL}). The power control policy used in the OTA aggregation  is the GBMA scheme in   \cite{sery2020analog}.
	\item{\bf Proposed 2}: In this method, the local device training and the server aggregation adopt the proposed scheme  in Algorithm (\ref{alg:genFWODL}). The power control policy used in the OTA aggregation  is the CI scheme in \cite{xing2021federated}.
	\item{\bf Baseline 1} (Nuclear norm \cite{yuan2021federated}): In each iteration, each local device updates the local model weights with the FeDMiD method in \cite{yuan2021federated} to solve  Problem (\ref{prob:nuclear}). Then the server  aggregates the local model weights by the FeDMiD algorithm in an OTA manner with the CI power policy and broadcasts the aggregated model weights to the local devices for them to execute their local calculation for the next iteration.
	\item{\bf Baseline 2} ($\ell_1$ regularization \cite{kumar2021pruning}): This method is a sparse model compression method. In each iteration, each local device will  update  the local model weights by optimizing the $\ell_1$-regularized objective  using one iteration of  Prox-SGD \cite{yang2019proxsgd}. Then, each device projects the sparse model weights  by the same  random matrix $\bm{A}\in\mathbb{R}^{M_{S}\times M}$  within each layer.  The results are then aggregated by the CI scheme at  the server, followed by an AMP \cite{donoho2009message} solver for decompression. 
		\item{\bf Baseline 3} (Magnitude pruning \cite{frankle2020linear}): This method is a sparse model compression method. In each iteration, each local device will  update  the local model weights by first optimizing the original problem 
using one iteration of  SGD and then keeping the top-$s$ weights with the largest magnitude. Then, it adopts the same communication method as Baseline 2.
		{     \item{\bf Baseline 4} (Uniform quantization \cite{UniformQ}): The element-wise uniform quantization is adapted to the model parameters in a layer-by-layer manner. Then the server aggregates the local model weights  in an OTA manner with the CI power policy and broadcasts the aggregated model weights to the local devices for them to execute their local calculation for the next iteration.}
\end{itemize}

\subsection{Experiment Settings and Preliminaries}\label{sec:expre}
In the  simulation experiments, we consider a deep-learning classification
task on the  MNIST  \cite{deng2012mnist} dataset, which is a  set of  handwritten digits, available from this page, has a training set of $60,000$ examples, and a test set of $10,000$ examples. Each example is a $28\times 28$ gray-scale image associated with a label from 10 classes. In the experiments, we use the first $50,000$ examples in the training set as the training data and the remaining examples in the training set as the test data.  The experiments  adopt the multilayer perceptrons (MLPs) model where the input dimension is $784$, the two  fully-connected layers have hidden nodes $\{256,256\}$, and the output dimension is $10$.   Overall, there are $268800$ weights and $522$ biases trained by   optimizing the {\em cross-entropy} loss with  the methods listed in Section \ref{sec:list}.  The training data samples are evenly and uniformly sampled and fed to the training methods with mini-batch size $D_m$. The simulation results are derived over $10$ Monte-Carlo trials with $K$ distributed IoT devices and a normalized noise variance of $\sigma_z^2=1$. {In the experiments, the collection of model weights of each layer is considered as one matrix variable and these model weight matrices are trained and aggregated in  a layer-by-layer manner.} {The compression ratio $\omega$ is defined as the ratio between the compressed size and the uncompressed size.

\subsection{Experimental Results and Discussion}

\subsubsection{Performance of the Proposed Scheme under Different  Design Parameters}\label{sec:percomp}
We first show the test accuracy of the proposed FedRLR scheme with a one standard deviation error band  under different mini-batch size $D_m$, device number $K$, and different penalty parameter $\mu$  in Fig. \ref{fig:diffdm}, Fig. \ref{fig:diffK}, and Fig. \ref{fig:diffdmu}{under the same compression ratio.} In the simulation, the transmit signal to noise ratio (SNR) is $\frac{P_{ave}}{\sigma^2_z}  = 25$ dB, where $P_{ave}$ is the averaged transmission power defined as $P_{ave}=\frac{\sum_{t=1}^{T_{iter}}\sum_{k=1}^KP_k^{(t)}}{KT_{iter}}$ with $P_k^{(t)}=(p^{(t)}_k)^2\|\bm{S}^{(t)}_{k}\|_F^2$. The learning rate is $\eta^{(t)}=\frac{q}{\nu+t}$, where we have $q=2$ and $\nu = 1000$. We can observe that during the training process, the test accuracy of the proposed scheme will keep increasing, and it will eventually converge to almost the same  test accuracy. Specifically, a larger mini-batch size $D_m$ or a larger $K$ indicates a smaller variance, leading to a slightly faster convergence speed.
{     Furthermore, compared to constant $\mu^{(t)}$, the proposed scheduling of  $\mu^{(t)}$, with $\mu^{(t)}=\frac{c_1}{\eta^{(t)}}$, will lead to better performance and the smaller the $c_1$ is, the faster the convergence will be, which is consistent with the convergence result in Lemma \ref{lem:stationary}.}
\begin{figure}[htbp]
\centering
	\includegraphics[width=0.7\linewidth]{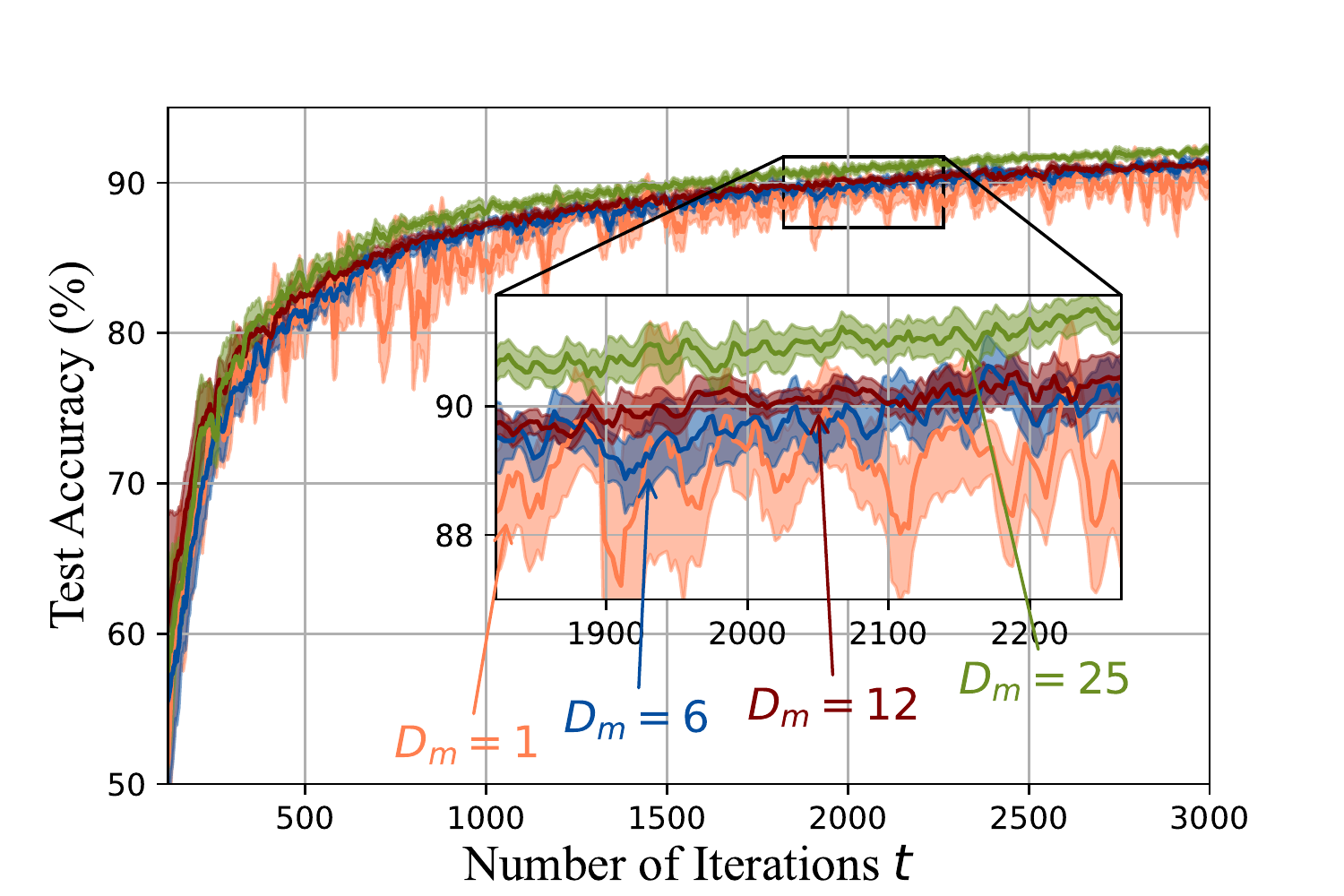}
	\caption{{Training performance of the proposed scheme under different mini-batch size $D_m$ with {$\omega=70\%$}, $\gamma=1$,  $K=10$ and $\mu^{(t)}=\frac{0.006}{\eta^{(t)}}$.}}
	\label{fig:diffdm}
\end{figure}
\begin{figure}[htbp]
\centering
	\includegraphics[width=0.7\linewidth]{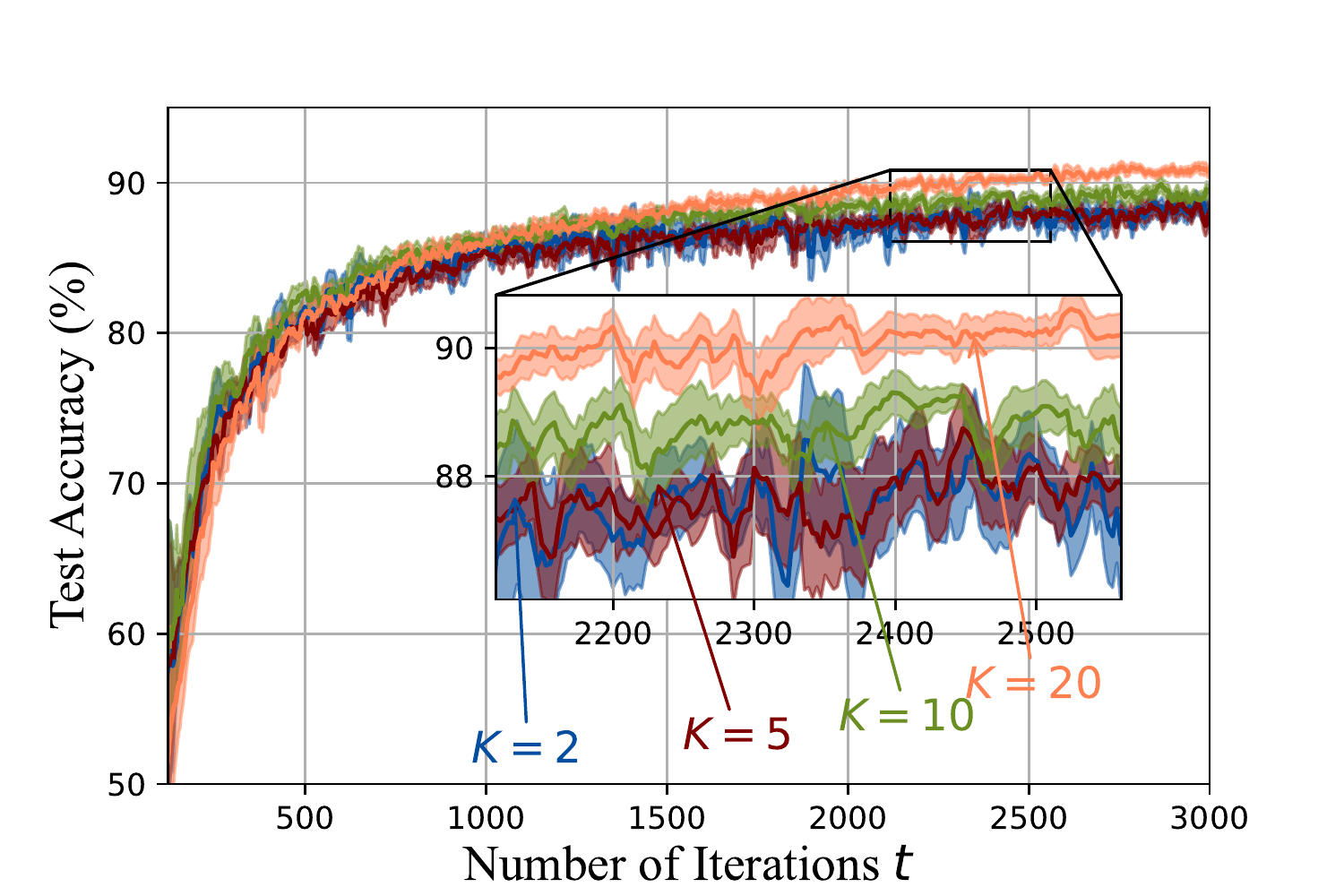}
	\caption{{ Training performance of the proposed scheme under different device number $K$ with {$\omega=70\%$}, $\gamma=1$,  $D_m=6$ and $\mu^{(t)}=\frac{0.006}{\eta^{(t)}}$.}}
	\label{fig:diffK}
\end{figure}
\begin{figure}[htbp]
\centering
	\includegraphics[width=0.7\linewidth]{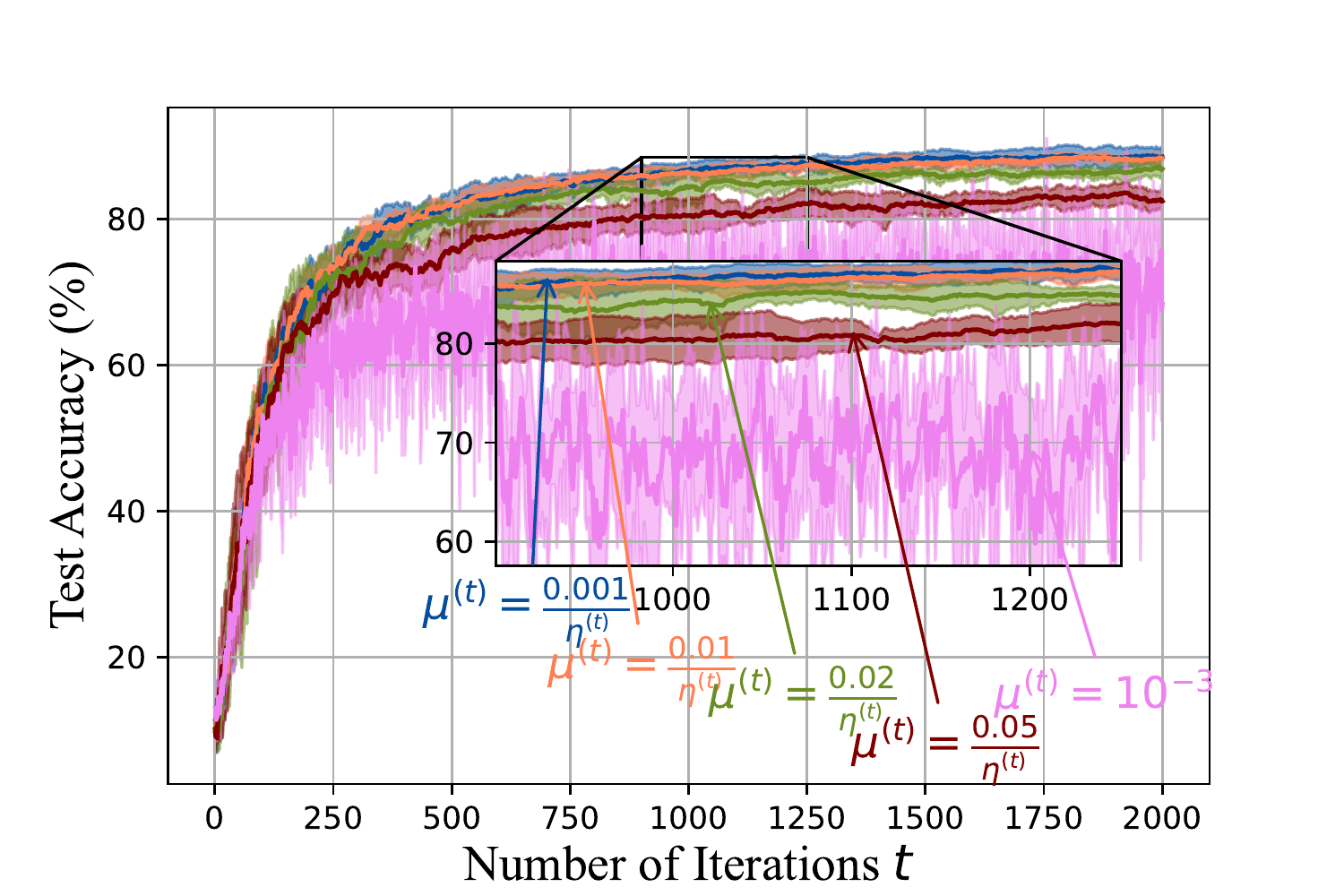}
	\caption{{       Training performance of the proposed scheme under different penalty parameter $\mu^{(t)}$ with {$\omega=70\%$}, $\gamma=1$,  $D_m=6$ and $K=10$.}}
	\label{fig:diffdmu}
\end{figure}

\subsubsection{Performance Comparison with Baselines}
We further compare the test accuracy   of the proposed scheme with existing
baselines for multi-class classification with  a one standard deviation error band,  under a fixed  transmission power budget throughout the training process, as shown in  Fig. \ref{fig:compTestacc}.  The transmit SNR is $\frac{P_{ave}}{\sigma^2_z}  = 25$ dB.   In the simulations, we set the mini-batch size $D_m=6$, the number of devices $K=10$, and the learning rate to $\eta^{(t)}=\frac{q}{\nu+t}$ with $q=2$ and $\nu=1000$ for all methods. In addition, we set $R=4$ in each layer and $\mu^{(t)}=\frac{0.006}{\eta^{(t)}}$ for the  proposed method.   The regularization parameters $\lambda$ in Baseline 1 and Baseline 2 are set as $0.0005$ and $0.02$, and the number of non-zero values in Baseline 3  occupies around $0.03125$ of the total number of the model weights.  The random projection matrix $\bm{A}\in\mathbb{R}^{8\times 256}$ in Baseline 2 and Baseline 3 is generated with elements that follow the standard normal distribution in an i.i.d manner for each layer.  We can observe
that  Baselines 2 and 3 have the worst performance mainly because  the AMP  decompression  requires that  the aggregated sparse model weights  from all devices  have  similar support, which is hard to satisfy in real applications. Therefore, there will be a considerably significant error in the AMP decompression process at the server.  Baseline 1 suffers from a higher effective noise level under the same transmit SNR, since the regularized nuclear norm objective can lead to a considerably large rank of the local model weights and the aggregated result is no longer low rank, which will further degrade the performance. In Baseline 4, there is no compression over the air. Therefore it can achieve better performance. The proposed algorithm performs better than all the other baselines and achieves almost the same performance as the benchmark.  {  The results of the proposed method with GBMA power control policy are better than that with the CI policy. This is due to the deep fading effect of the channel which will lead to a higher effective noise level for CI power policy under the same transmit power budget.}

\begin{figure}[htbp]
	\centering
	\includegraphics[width=0.7\linewidth]{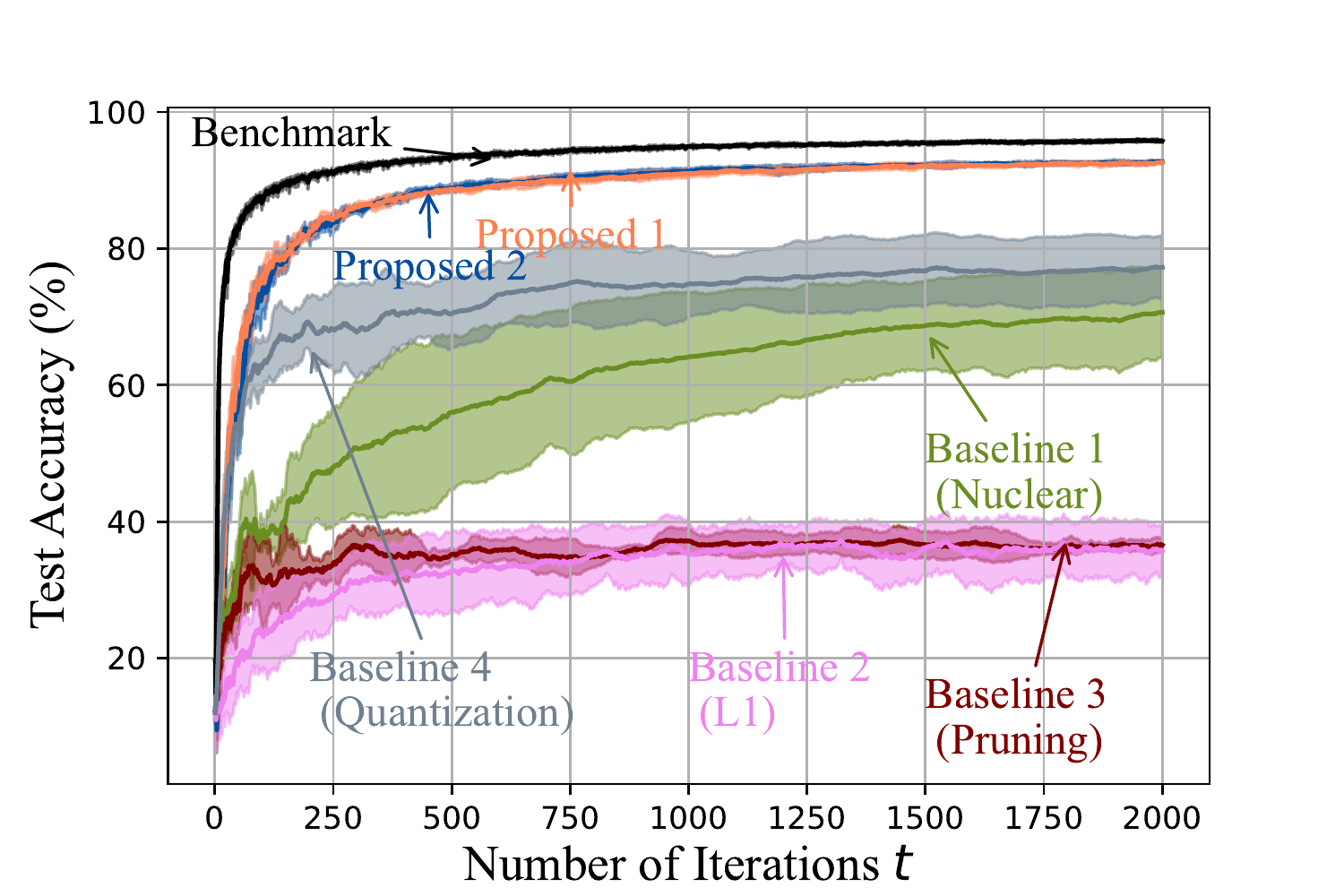}
	\caption{{ Performance comparison  of the proposed scheme with existing
			baselines for multi-class classification on the Fashion MNIST dataset under the same compression ratio $\omega=70\%$.}}\label{fig:compTestacc}
\end{figure}

{Fig.\ref{fig:DiffCR} shows the accuracy  comparison under different compression ratios with a one standard deviation error bar, where we set the SNR to  $25$ dB.  The mini-batch size $D_m$ is $6$, the number of devices $K$ is $10$, and the learning rate is $\eta^{(t)}=\frac{q}{\nu+t}$ with $q=2$ and $\nu=1000$ for all methods. In addition, we set  $\mu^{(t)}=\frac{0.006}{\eta^{(t)}}$ for the  proposed method.  The rank $R$ for Proposed 1 and Proposed 2, the regularization parameter $\lambda$ for Baseline 1 and Baseline 2, and the number of non-zero values  in Baseline 3  is adjusted to achieve the desired compression ratio. The random projection matrix $\bm{A}\in\mathbb{R}^{ \frac{ 256}{\omega}\times 256}$ in Baseline 2 and Baseline 3 is generated with elements that follow the standard normal distribution in an i.i.d manner for each layer.   The results in Fig.\ref{fig:DiffCR}  indicate that the proposed FedRLR achieves a better performance under various compression ratios. Furthermore, it is less sensitive to high compression ratios compared to the baselines with an  accuracy  higher than $80\%$ under all compression ratios in the simulation. 
\begin{figure}[htbp]
\centering
	\includegraphics[width=0.7\linewidth]{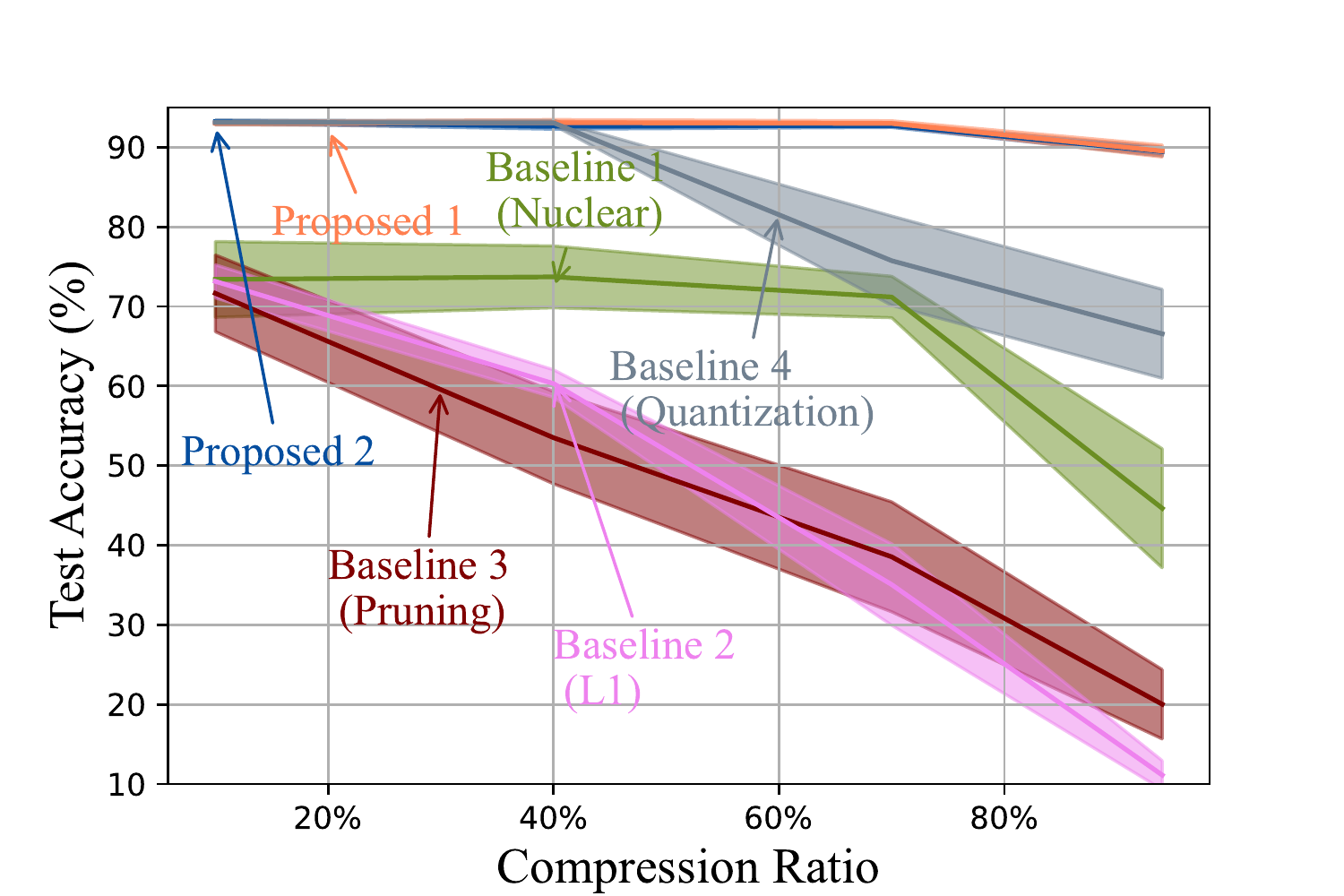}
	\caption{ {Performance comparison  of the proposed scheme with existing
			baselines for multi-class classification on the Fashion MNIST dataset  under different compression ratios $\omega$.}}
	\label{fig:DiffCR}
\end{figure}

Table \ref{tab:1} lists  the communication overhead required for each method to achieve  $70\%$ accuracy with the benchmark method as a reference under the same settings as those in Fig.\ref{fig:compTestacc}. Since Baseline 2 and 3 can not achieve $70\%$ accuracy under these settings, we  set the random projection matrix $\bm{A}$ with dimension $200\times 256$ to guarantee a better performance of Baseline 2 and 3. As we can see, all the model compression methods  can save the communication overhead compared to the benchmark thanks to the OTA aggregation. In the proposed method,  the factorized model transmission with OTA aggregation helps to achieve a much lower communication overhead compared to the benchmark. Furthermore, the efficient alternating optimization over the Riemannian manifold leads to a faster convergence speed and enables  the proposed FedRLR to achieve  $70\%$ accuracy with less communication overhead compared to the baselines.
\begin{savenotes}
 \begin{table}[htbp]
	\centering
	\begin{threeparttable}
		\caption{Comparison of Communication Overhead to Achieve $70\%$ Accuracy}\label{tab:1}
		\begin{tabular}{cccc}
			\toprule
			Methods & Communication Overhead\\
			\hline
		Benchmark\footnote{Here the communication overhead for the benchmark is regarded as the reference value. The absolute communication overhead for the benchmark to achieve $70\%$ accuracy is $2.02\times 10^7$ sub-carriers if one real-valued weight is transmitted by one sub-carrier. }& $100\%$\\
			\rowcolor{light-gray}
			Proposed 1& $1.42\%$\\
			Proposed 2& $1.44\%$\\
			\rowcolor{light-gray}
			Baseline 1 (Nuclear norm)& $87.23\%$\\
			Baseline 2 ($\ell_1$ regularization) & $75.49\%$\\
				\rowcolor{light-gray}
			Baseline 3 (Magnitude pruning) & $93.13\%$\\
   Baseline 4 (Quantization) & $100\%$\\
			\bottomrule
		\end{tabular}
	\end{threeparttable}
\end{table}
\end{savenotes}}
\subsubsection{ Wall Clock Time Comparison with Baselines}
In this subsection, we  show the complexity comparison  in terms of the  wall clock CPU time during inference in the following Fig. \ref{fig:comptime}.
\begin{figure}[htbp]
\centering
	\includegraphics[width=0.7\linewidth]{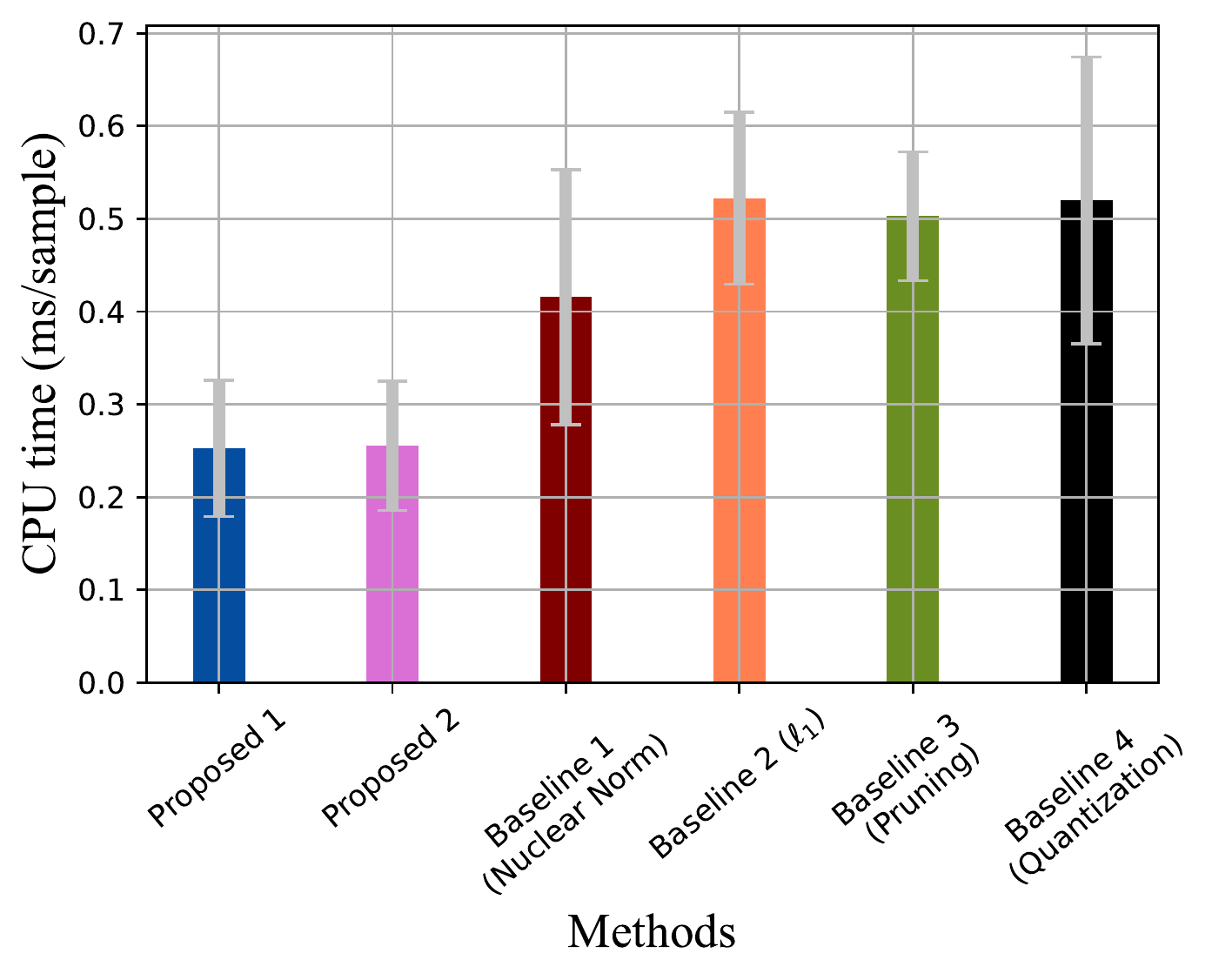}
	\caption{{Wall clock
 inference time  to achieve  the same accuracy at $70\%$.}}
	\label{fig:comptime}
\end{figure}

As we can see, the proposed low-rank model compression achieves the fastest time for inference thanks to the optimization over the Riemannian manifold that guarantees the low-rankness of the produced model. The wall clock time comparison  demonstrates the efficiency of the proposed federated  low-rank model compression, FedRLR, to reduce the computation load.

\section{Conclusion}\label{sec:con}
In this paper, we  proposed the efficient  FedRLR  model compression scheme with OTA aggregation in the FL systems, which leverages the benign properties of the fixed-rank Riemannian manifold to enable the exact low-rankness of the local model weights and the compatibility of the OTA aggregation. The FedRLR scheme directly optimizes the local and global model weights over the Riemannian  manifold in an alternating manner, thereby preserving  the  low-rankness of the local and global model weights as well as protecting the convergence trajectory after the OTA aggregation. Based on the proposed
algorithm, we established the convergence analysis. Simulation results illustrated that the
proposed algorithm can achieve a much faster convergence
rate and lower training loss than existing baselines. {      Although our proposed scheme is specifically designed for OTA communication in federated learning, it can still provide performance advantages in point-to-point links or other digital FL schemes where co-channel interference is not present. For example, The proposed method can guarantee a low-rank model without tuning hyperparameters  and the low-rank model can save the computation cost for inference. In future work, the proposed scheme together with the convergence analysis can be extended into more general conditions. }

\appendices
\section{Derivation of the Riemannian gradient}\label{der:rg}
Since $\mathcal{M}_R$ is an embedded submanifold in $\mathbb{R}^{M\times N}$, the Riemannian gradient is given as the orthogonal projection onto the tangent space of the Euclidean gradient of 
$\psi(\bm{\Theta}_k)=\frac{1}{DK}\sum_{l=1}^Df(\bm{\Theta}_k,\bm{\zeta}_{k,l})+\frac{\mu}{2K}\|\bm{\Theta}_{0}^{(t)}-\bm{\Theta}_{k}\|_F^2 $ seen as a function on $\mathbb{R}^{M\times N}$  \cite{absil2009optimization}. Based on the tangent space defined in defined in \cite{wei2016guarantees},  the orthogonal projection onto the tangent space of $\mathcal{M}_R$ at $\bm{\Theta}_{k}^{(t)}$ is
{\small
\begin{equation}\label{eq:projtan}
\begin{aligned}
&P_{\mathcal{T}_{\bm{\Theta}_k^{(t)}}\mathcal{M}_R}:\mathbb{R}^{M\times N}\mapsto\mathcal{T}_{\bm{\Theta}_k^{(t)}}\mathcal{M}_R,\\
&\bm{\Delta}\mapsto \\
&\bm{U}^{(t)}_k(\bm{U}^{(t)}_k)^T\bm{\Delta}+\bm{\Delta}\bm{V}^{(t)}_k(\bm{V}^{(t)}_k)^T-\bm{U}^{(t)}_k(\bm{U}^{(t)}_k)^T\bm{\Delta}\bm{V}^{(t)}_k(\bm{V}^{(t)}_k)^T,
\end{aligned}
\end{equation}}
where $\bm{U}^{(t)}_k$ and $\bm{V}^{(t)}_k$ are obtained by the compact SVD of $\bm{\Theta}_k^{(t)}$  via \begin{equation}
    SVD_{comp}(\bm{\Theta}_k^{(t)})=[\bm{U}^{(t)}_k,\bm{\Sigma}^{(t)}_k,\bm{V}^{(t)}_k].
\end{equation} 
Since the Euclidean gradient of 
$\psi(\bm{\Theta}_k)$ is 
{\small\begin{equation}\label{eq:eucgrad}
\begin{aligned}
\nabla_{\bm{\Theta}_k}\psi(\bm{\Theta}_k)=
\nabla_{\bm{\Theta}_k} \frac{1}{KD}\sum_{l=1}^Df(\bm{\Theta}^{(t)}_k,\bm{\zeta}_{k,l})+\frac{\mu}{K}\bm{\Theta}^{(t)}_k-\frac{\mu}{K}\bm{\Theta}^{(t)}_0.
\end{aligned}
\end{equation}}
Combining \ref{eq:projtan} and \ref{eq:eucgrad}, we have the Riemannian gradient $\bm{g}_k^{(t)}=P_{\mathcal{T}_{\bm{\Theta}_k^{(t)}}\mathcal{M}_R}(\nabla_{\bm{\Theta}_k}\psi(\bm{\Theta}_k))$, which finishes the derivation.

\section{Proof of Lemma \ref{lem:fixlip}}\label{proof:lem2}
Let $ \mathcal{N}_{\bm{\Theta}_k}\mathcal{M}_R$ be the  normal space of $\mathcal{M}_R$ at ${\bm{\Theta}_k}=\bm{U}_k^{(t)}\bm{\Sigma}_k^{(t)}(\bm{V}_k^{(t)})^T$. Then we have
{\small\begin{equation}
\begin{aligned}
   &\mathcal{P}_{\mathcal{N}_{\bm{\Theta}_k}\mathcal{M}_R}(\bm{Z})\\
   =&\big(\bm{I}_M-\bm{U}_k^{(t)}(\bm{U}_k^{(t)})^T\big)\big(\bm{Z}\big)\big(\bm{I}_N-\bm{V}_k^{(t)}(\bm{V}_k^{(t)})^T\big),
\end{aligned}
\end{equation}}
where $\bm{Z}\in\mathbb{R}^{M\times N}$.

Since $$grad_{\bm{\Theta}_k}\frac{1}{D}\sum_{l=1}^Df(\bm{\Theta}_k,\bm{\zeta}_{k,l})=\mathcal{P}_{\mathcal{T}_{\bm{\Theta}_k}\mathcal{M}_R}\big(\nabla_{\bm{\Theta}_k}\frac{1}{D}\sum_{l=1}^Df(\bm{\Theta}_k,\bm{\zeta}_{k,l})\big),$$ we have
{\small\begin{equation}\label{eq:gradnabla}
    \begin{aligned}
       &\langle grad_{\bm{\Theta}_k}\frac{1}{D}\sum_{l=1}^Df(\bm{\Theta}_k,\bm{\zeta}_{k,l}),\bm{\Theta}^{'}_k- \bm{\Theta}_k\rangle\\
       =& \langle \nabla_{\bm{\Theta}_k}\frac{1}{D}\sum_{l=1}^Df(\bm{\Theta}_k,\bm{\zeta}_{k,l}),\mathcal{P}_{\mathcal{T}_{\bm{\Theta}_k}\mathcal{M}_R}\big(\bm{\Theta}^{'}_k- \bm{\Theta}_k\big)\rangle\\
       \leq&\langle \nabla_{\bm{\Theta}_k}\frac{1}{D}\sum_{l=1}^Df(\bm{\Theta}_k,\bm{\zeta}_{k,l}),\big(\bm{\Theta}^{'}_k- \bm{\Theta}_k\big)\rangle\\
       &+3\| \nabla_{\bm{\Theta}_k}\frac{1}{D}\sum_{l=1}^Df(\bm{\Theta}_k,\bm{\zeta}_{k,l})\|_2\|\bm{\Theta}^{'}_k- \bm{\Theta}_k\|^2_2\\
       \leq&\langle \nabla_{\bm{\Theta}_k}\frac{1}{D}\sum_{l=1}^Df(\bm{\Theta}_k,\bm{\zeta}_{k,l}),\big(\bm{\Theta}^{'}_k- \bm{\Theta}_k\big)\rangle\\
       &+3B_1\|\bm{\Theta}^{'}_k- \bm{\Theta}_k\|^2_2.
    \end{aligned}
\end{equation}}
Therefore, combining Assumption 1-2) and (\ref{eq:gradnabla}), we have 
{\small\begin{equation}
    \begin{aligned}
    &\Bigg|\frac{1}{D}\sum_{l=1}^Df(\bm{\Theta}_k,\bm{\zeta}_{k,l})-\bigg[\frac{1}{D}\sum_{l=1}^Df(\bm{\Theta}^{'}_k,\bm{\zeta}_{k,l})\\
    &+\langle grad_{\bm{\Theta}_k} \frac{1}{D}\sum_{l=1}^Df(\bm{\Theta}^{'}_k,\bm{\zeta}_{k,l}),\bm{\Theta}_k-\bm{\Theta}^{'}_k\rangle\bigg]\Bigg|\\
    &\leq \frac{L+6B_1}{2}\|\bm{\Theta}_k-\bm{\Theta}^{'}_k\|_F^2.
    \end{aligned}
    \end{equation}}
\section{Proof of Lemma \ref{lem:convconsen}}\label{proof:lem3}
According to the definitions of $\vec{\bm{\Theta}}$ and $\vec{\bm{\Theta}_0^*}$, we have
{\small\begin{equation}
    \begin{aligned}
       &\mathbb{E}_t[\|\vec{\bm{\Theta}}^{(t+1)}-\vec{\bm{\Theta}_0^*}^{(t+1)}\|_F^2]\leq\mathbb{E}_t[\|\vec{\bm{\Theta}}^{(t+1)}-\vec{\bm{\Theta}_0^*}^{(t)}\|_F^2]\\
       =&\sum_{k=1}^K\mathbb{E}_t[\|\mathcal{R}_{\bm{\Theta}_k^{(t)}}(-\eta^{(t)}\bm{\hat g}^{(t)}_k)-(\bm{\Theta}_0^*)^{(t)}\|_F^2\\
           \end{aligned}
\end{equation}}
{\small\begin{equation}
    \begin{aligned}
       \overset{(a)}\leq&\sum_{k=1}^K\mathbb{E}_t[\|\bm{\Theta}_k^{(t)}-\eta^{(t)}\bm{\hat g}^{(t)}_k-(\bm{\Theta}_0^*)^{(t)}\|_F^2\\
         \end{aligned}
\end{equation}}
{\small\begin{equation}
    \begin{aligned}
       \overset{(b)}\leq&\sum_{k=1}^K\mathbb{E}_t\bigg[\Big\|\bm{\Theta}_k^{(t)}-(\bm{\Theta}_0^*)^{(t)}-\eta^{(t)}\frac{\mu}{K}(\bm{U}^{(t)}_k(\bm{U}^{(t)}_k)^T\Big(\bm{\Theta}^{(t)}_k-\bm{\Theta}^{(t)}_0\Big)\\
       &+\Big(\bm{\Theta}^{(t)}_k-\bm{\Theta}^{(t)}_0\Big)\bm{V}^{(t)}_k(\bm{V}^{(t)}_k)^T\\
       &-\bm{U}^{(t)}_k(\bm{U}^{(t)}_k)^T\Big(\bm{\Theta}^{(t)}_k-\bm{\Theta}^{(t)}_0\Big)\bm{V}^{(t)}_k(\bm{V}^{(t)}_k)^T)\Big\|_F^2\bigg]\\
       &+(\eta^{(t)})^2KB_1^2\\
         \end{aligned}
\end{equation}}
{\small\begin{equation}
    \begin{aligned}
       \overset{(c)}\leq& \sum_{k=1}^K\bigg[(1-\eta^{(t)}\frac{\mu}{K})^2\|\bm{\Theta}_k^{(t)}-(\bm{\Theta}_0^*)^{(t)}\|_F^2\bigg]+(\eta^{(t)})^2KB_1^2\\
       =&(1-\eta^{(t)}\frac{\mu^{(t)}}{K})^2\|\vec{\bm{\Theta}}^{(t)}-\vec{\bm{\Theta}_0^*}^{(t)}\|_F^2+(\eta^{(t)})^2KB_1^2,
    \end{aligned}
\end{equation}}
where $(a)$ is obtained in a similar way to \cite[Lemma 2]{li2021weakly}   by the fact that
projections onto closed convex sets are 1-Lipschitz, $(b)$ is from the Assumption 1-1), and $(c)$ is from the unbiased OTA aggregation. We represent the time-variant penalty parameter $\mu$ by $\mu^{(t)}$ in the last equality.
Take expectation over all randomness, we have
{\small\begin{equation}
    \begin{aligned}
       &\mathbb{E}[\|\vec{\bm{\Theta}}^{(t+1)}-\vec{\bm{\Theta}_0^*}^{(t+1)}\|_F^2]\\\leq&(1-\eta^{(t)}\frac{\mu^{(t)}}{K})^2\mathbb{E}[\|\vec{\bm{\Theta}}^{(t)}-\vec{\bm{\Theta}_0^*}^{(t)}\|_F^2]+(\eta^{(t)})^2KB_1^2\\
       =&(1-\frac{c_1}{K})^2\mathbb{E}[\|\vec{\bm{\Theta}}^{(t)}-\vec{\bm{\Theta}_0^*}^{(t)}\|_F^2]+(\eta^{(t)})^2KB_1^2.
    \end{aligned}
\end{equation}}
Let $\alpha^{(t)}=\frac{\mathbb{E}[\|\vec{\bm{\Theta}}^{(t)}-\vec{\bm{\Theta}_0^*}^{(t)}\|_F^2]}{K(\eta^{(t)})^2}$, then we have
{\small\begin{equation}
    \begin{aligned}
       &\alpha^{(t+1)}\\
       \leq&(1-\frac{c_1}{K})^2\alpha^{(t+1)}+\frac{(\eta^{(t)})^2}{\eta^{(t+1)})^2}B_1^2\\
       =&\big((1-\frac{c_1}{K})^2\big)^{(t+1-T_c)}\alpha^{(T_c)}+B_1^2\sum_{t'=T_c}^t
    \big((1-\frac{c_1}{K})^2\big)^{(t-t')}\frac{(\eta^{(t')})^2}{\eta^{(t'+1)})^2},\end{aligned}
\end{equation}}
with some $T_c$ such that $\frac{(\eta^{(t')})^2}{\eta^{(t'+1)})^2}\leq 2, \forall t'>T_c$. According to \cite [Proposition 8]{liu2017convergence}, we finish the proof.

\section{Proof of Lemma \ref{lem:stationary}}\label{proof:lem4}
Let $F(\vec{\bm{\Theta}})=\frac{1}{K}\sum_{k=1}^K\frac{1}{D}\sum_{l=1}^Df(\bm{\Theta}_k,\bm{\zeta}_{k,l})$ and $\bm{\hat{g}}_1^{(t)}=\bigg[(\bm{\hat{g}}_{1,1}^{(t)})^T,(\bm{\hat{g}}_{2,1}^{(t)})^T,\ldots,(\bm{\hat{g}}_{K,1}^{(t)})^T\bigg]^T$, where we have $\bm{\hat{g}}_{k,1}^{(t)}=grad_{\bm{\Theta}_k} \frac{1}{K|\mathcal{D}^{(t)}_k|}\sum_{\bm{\zeta}_{k,l}\in\mathcal{D}^{(t)}_k}f(\bm{\Theta}^{(t)}_k,\bm{\zeta}_{k,l})$. Due to Assumption 1-3), we have $\mathbb{E}[\bm{\hat{g}}_{1}^{(t)}]= grad_{\vec{\bm{\Theta}}}F(\vec{\bm{\Theta}}^{(t)})$ and $\mathbb{E}[\bm{\hat{g}}_{k,1}^{(t)}]= grad_{\vec{\bm{\Theta}_k}}F(\vec{\bm{\Theta}}^{(t)})$ . Then we have the following inequality.
{\small\begin{equation}\label{eq:sta1}
\begin{aligned}
&\mathbb{E}_{t}[F(\vec{\bm{\Theta}}^{(t+1)})]\\
\leq& F(\vec{\bm{\Theta}}^{(t)})+\langle grad_{\vec{\bm{\Theta}}}F(\vec{\bm{\Theta}}^{(t)}),\mathbb{E}_{t}[\vec{\bm{\Theta}}^{(t+1)}-\vec{\bm{\Theta}}^{(t)}]\rangle\\
&+\frac{L+6B_1}{2}\mathbb{E}_{t}[\|\vec{\bm{\Theta}}^{(t+1)}-\vec{\bm{\Theta}}^{(t)}\|_F^2]\quad (\text{Lemma 2})\\
=&F(\vec{\bm{\Theta}}^{(t)})-\langle grad_{\vec{\bm{\Theta}}}F(\vec{\bm{\Theta}}^{(t)}),\mathbb{E}_{t}[\eta^{(t)}\bm{\hat{g}}_1^{(t)}]\rangle\\
&+\langle grad_{\vec{\bm{\Theta}}}F(\vec{\bm{\Theta}}^{(t)}),\mathbb{E}_{t}[\vec{\bm{\Theta}}^{(t+1)}-\vec{\bm{\Theta}}^{(t)}+\eta^{(t)}\bm{\hat{g}}_1^{(t)}]\rangle \\
&+\frac{L+6B_1}{2}\mathbb{E}_{t}[\|\vec{\bm{\Theta}}^{(t+1)}-\vec{\bm{\Theta}}^{(t)}\|_F^2\\
=&F(\vec{\bm{\Theta}}^{(t)})- \eta^{(t)}\|grad_{\vec{\bm{\Theta}}}F(\vec{\bm{\Theta}}^{(t)})\|_F^2\\
&+\langle grad_{\vec{\bm{\Theta}}}F(\vec{\bm{\Theta}}^{(t)}),\mathbb{E}_{t}[\vec{\bm{\Theta}}^{(t+1)}-\vec{\bm{\Theta}}^{(t)}+\eta^{(t)}\bm{\hat{g}}_1^{(t)}]\rangle \\
&+\frac{L+6B_1}{2}\mathbb{E}_{j}[\|\vec{\bm{\Theta}}^{(j+1)}-\vec{\bm{\Theta}}^{(t)}\|_F^2]\quad(\text{Assumption 1-3)})
\end{aligned}
\end{equation}}
{\small\begin{equation}\label{eq:sta2}
\begin{aligned}
(\ref{eq:sta1})\leq& F(\vec{\bm{\Theta}}^{(t)})- \eta^{(t)}\|grad_{\vec{\bm{\Theta}}}F(\vec{\bm{\Theta}}^{(t)})\|_F^2+\frac{1}{2\epsilon}\|grad_{\vec{\bm{\Theta}}}F(\vec{\bm{\Theta}}^{(t)})\|_F^2\\
&+\frac{\epsilon}{2} \mathbb{E}_{t}[\|\vec{\bm{\Theta}}^{(t+1)}-\vec{\bm{\Theta}}^{(t)}+\eta^{(t)}\bm{\hat{g}}_1^{(t)}\|_F^2] \\&+\frac{L+6B_1}{2}\mathbb{E}_{t}[\|\vec{\bm{\Theta}}^{(t+1)}-\vec{\bm{\Theta}}^{(t)}\|^2_F]\\
&\quad\quad(\text{Young's inequality and Jensen's inequality})\\
\leq& F(\vec{\bm{\Theta}}^{(t)})- \eta^{(t)}\|grad_{\vec{\bm{\Theta}}}F(\vec{\bm{\Theta}}^{(t)})\|_F^2+\frac{1}{2\epsilon}\|grad_{\vec{\bm{\Theta}}}F(\vec{\bm{\Theta}}^{(t)})\|_F^2\\
&+\frac{\epsilon}{2} \mathbb{E}_{t}[2\|\vec{\bm{\Theta}}^{(t+1)}-\vec{\bm{\Theta}}^{(t)}\|_F^2+2(\eta^{(t)})^2\|\bm{\hat{g}}_1^{(t)}\|_F^2] \\&+\frac{L+6B_1}{2}\mathbb{E}_{t}[\|\vec{\bm{\Theta}}^{(t+1)}-\vec{\bm{\Theta}}^{(t)}\|^2_F]\quad(\text{Cauchy's inequality})\\
\leq& F(\vec{\bm{\Theta}}^{(t)})- (\eta^{(t)}-\epsilon(\eta^{(t)})^2-\frac{1}{2\epsilon})\|grad_{\vec{\bm{\Theta}}}F(\vec{\bm{\Theta}}^{(t)})\|_F^2\\
&+\frac{L+6B_1+2\epsilon}{2}\mathbb{E}_{t}\Bigg[\sum_{k=1}^K\bigg\|\eta^{(t)}\bm{\hat g}^{(t)}_{k,1}\\
&\quad+grad_{\bm{\Theta}_{k}}\frac{c_1}{2K}\|\bm{\Theta}^{(t)}_0-\bm{\Theta}^{(t)}_{k}\|_F^2\bigg\|^2_F\Bigg]\\
&\quad\quad(\text{Projection-like Retraction})\\
\end{aligned}
\end{equation}}
{\small\begin{equation}\label{eq:sta3}
\begin{aligned}
(\ref{eq:sta2})\leq& F(\vec{\bm{\Theta}}^{(t)})- (\eta^{(t)}-\epsilon(\eta^{(t)})^2-\frac{1}{2\epsilon})\|grad_{\vec{\bm{\Theta}}}F(\vec{\bm{\Theta}}^{(t)})\|_F^2\\
&+\frac{L+6B_1+2\epsilon}{2}2(\eta^{(t)})^2\|grad_{\vec{\bm{\Theta}}}F(\vec{\bm{\Theta}}^{(t)})\|_F^2\\
&+2(\eta^{(t)})^2\frac{\sigma^2_g}{D_m}
+2\frac{c^2_1}{K^2}\|\vec{\bm{\Theta}}^{(t)}-\vec{\bm{\Theta}_0^*}^{(t)}\|_F^2.\\
&\quad\quad(\text{Assumption 1-4) and property  of variance})
\end{aligned}
\end{equation}}
Combining (\ref{eq:sta3}) with Lemma 3, we have
{\small\begin{equation}\label{eq:sta4}
\begin{aligned}
&\mathbb{E}_{t}[F(\vec{\bm{\Theta}}^{(t+1)})]\\
\leq& F(\vec{\bm{\Theta}}^{(t)})- (\eta^{(t)}-\epsilon(\eta^{(t)})^2-\frac{1}{2\epsilon})\|grad_{\vec{\bm{\Theta}}}F(\vec{\bm{\Theta}}^{(t)})\|_F^2\\
&+({L+6B_1+2\epsilon})(\eta^{(t)})^2\|grad_{\vec{\bm{\Theta}}}F(\vec{\bm{\Theta}}^{(t)})\|_F^2\\
&+2(\eta^{(t)})^2K\frac{\sigma^2_g}{D_m}
+2\frac{c^2_1c_2B^2_1(\eta^{(t)})^2\sigma^2_z}{\gamma K}.\\
\end{aligned}
\end{equation}}
Taking expectation with respect to all the randomness and rearranging (\ref{eq:sta4}), we have
{\small\begin{equation}\label{eq:sta5}
\begin{aligned}
&\inf_{1\leq t\leq T_{iter}}\mathbb{E}[\|grad_{\vec{\bm{\Theta}}}F(\vec{\bm{\Theta}}^{(t)})\|_F^2]\\
\leq&\frac{C^*+\sum_{t=0}^{T_{iter}}2(\eta^{(t)})^2\frac{\sigma^2_g}{D_m}
+2\frac{c^2_1c_2B^2_1(\eta^{(t)})^2\sigma^2_z}{\gamma K}}{\sum_{t=0}^{T_{iter}}\eta^{(t)}-{(L+6B_1+3\epsilon)}(\eta^{(t)})^2-\frac{1}{2\epsilon}},
\end{aligned}
\end{equation}}
when $\epsilon\geq\frac{2L+6B_1}{3}$.
Choosing $\bm{\Upsilon}_k=\bm{0},\forall k$, we finish the proof.

\section{Proof of Theorem \ref{thm:converKKT}}\label{proof:thm1}
Since the numerator of (\ref{eq:convrate}) is at the order of $\sum^{T_{iter}}_{t=0}\mathcal{O}(t^{-2})\leq \infty$ and the denominator is at the order of $\sum^{T_{iter}}_{t=0}\mathcal{O}(t^{-1})\to \infty$ when $T_{iter}\to \infty$, we have 
{\small\begin{equation}
       \underset{t\to \infty}{\lim} \bm{\Theta}^{(t)}_k\to(\bm{\Theta}^*_0)^{(t)}, \quad  \bm{\Theta}^{(t)}_k \in \mathcal{M}_R, \forall k\in[K], (\bm{\Theta}^*_0)^{(t)} \in \mathcal{M}_R,
    \end{equation}
    \begin{equation}
    \begin{aligned}
    & \underset{t\to \infty}{\lim} grad_{\vec{\bm{\Theta}}}L_{\mathscr{P}2}(\vec{\bm{\Theta}^{(t)}},\bm{\Theta}^{(t)}_0,\vec{\bm{\Upsilon}})\to \bm{0},
    \end{aligned}
\end{equation}}
and
{\small\begin{equation}
    \begin{aligned}
 &\underset{t\to \infty}{\lim}grad_{{\bm{\Theta}^{(t)}_0}}L_{\mathscr{P}2}(\vec{\bm{\Theta}^{(t)}},\bm{\Theta}^{(t)}_0,\vec{\bm{\Upsilon}})\to \bm{0}.
     \end{aligned}
\end{equation}}
    in expectation.

{\footnotesize
	\bibliography{IEEEabrv,Reference}}

\begin{thebibliography}{10}
\providecommand{\url}[1]{#1}
\csname url@samestyle\endcsname
\providecommand{\newblock}{\relax}
\providecommand{\bibinfo}[2]{#2}
\providecommand{\BIBentrySTDinterwordspacing}{\spaceskip=0pt\relax}
\providecommand{\BIBentryALTinterwordstretchfactor}{4}
\providecommand{\BIBentryALTinterwordspacing}{\spaceskip=\fontdimen2\font plus
\BIBentryALTinterwordstretchfactor\fontdimen3\font minus
  \fontdimen4\font\relax}
\providecommand{\BIBforeignlanguage}[2]{{%
\expandafter\ifx\csname l@#1\endcsname\relax
\typeout{** WARNING: IEEEtran.bst: No hyphenation pattern has been}%
\typeout{** loaded for the language `#1'. Using the pattern for}%
\typeout{** the default language instead.}%
\else
\language=\csname l@#1\endcsname
\fi
#2}}
\providecommand{\BIBdecl}{\relax}
\BIBdecl

\bibitem{Simonyan15}
K.~Simonyan and A.~Zisserman, ``Very deep convolutional networks for
  large-scale image recognition,'' in \emph{Inter. Conf. Learn.
  Representations}, 2015.

\bibitem{NEURIPS2020_1457c0d6}
T.~Brown \emph{et~al.}, ``Language models are few-shot learners,'' in
  \emph{NeurIPS}, vol.~33.\hskip 1em plus 0.5em minus 0.4em\relax Curran
  Associates, Inc., 2020, pp. 1877--1901.

\bibitem{gong2014compressing}
Y.~Gong, L.~Liu, M.~Yang, and L.~Bourdev, ``Compressing deep convolutional
  networks using vector quantization,'' \emph{arXiv preprint arXiv:1412.6115},
  2014.

\bibitem{wu2016quantized}
J.~Wu, C.~Leng, Y.~Wang, Q.~Hu, and J.~Cheng, ``Quantized convolutional neural
  networks for mobile devices,'' in \emph{IEEE Conf. Comput. Vis. Pattern
  Recog.}, 2016, pp. 4820--4828.

\bibitem{frankle2020linear}
J.~Frankle, G.~K. Dziugaite, D.~Roy, and M.~Carbin, ``Linear mode connectivity
  and the lottery ticket hypothesis,'' in \emph{Inter. Conf. Machine
  Learn.}\hskip 1em plus 0.5em minus 0.4em\relax PMLR, 2020, pp. 3259--3269.

\bibitem{han2015learning}
S.~Han, J.~Pool, J.~Tran, and W.~Dally, ``Learning both weights and connections
  for efficient neural network,'' \emph{NeurIPS}, vol.~28, 2015.

\bibitem{lebedev2016fast}
V.~Lebedev and V.~Lempitsky, ``Fast convnets using group-wise brain damage,''
  in \emph{IEEE Conf. Comput. Vis. Pattern Recog.}, 2016, pp. 2554--2564.

\bibitem{kumar2021pruning}
A.~Kumar, A.~M. Shaikh, Y.~Li, H.~Bilal, and B.~Yin, ``Pruning filters with
  l1-norm and capped l1-norm for cnn compression,'' \emph{Applied
  Intelligence}, vol.~51, no.~2, pp. 1152--1160, 2021.

\bibitem{vogels2019powersgd}
T.~Vogels, S.~P. Karimireddy, and M.~Jaggi, ``Power{SGD}: Practical low-rank
  gradient compression for distributed optimization,'' \emph{arXiv preprint
  arXiv:1905.13727}, 2019.

\bibitem{cho2019gradzip}
M.~Cho, V.~Muthusamy, B.~Nemanich, and R.~Puri, ``Gradzip: Gradient compression
  using alternating matrix factorization for large-scale deep learning,'' in
  \emph{NeurIPS}, 2019.

\bibitem{lowrankrelax}
J.~M. Alvarez and M.~Salzmann, ``Compression-aware training of deep networks,''
  in \emph{Proceedings of the 31st International Conference on Neural
  Information Processing Systems}, ser. NIPS'17.\hskip 1em plus 0.5em minus
  0.4em\relax Red Hook, NY, USA: Curran Associates Inc., 2017, p. 856–867.

\bibitem{yuan2021federated}
H.~Yuan, M.~Zaheer, and S.~Reddi, ``Federated composite optimization,'' in
  \emph{Inter. Conf. on Machine Learn.}\hskip 1em plus 0.5em minus 0.4em\relax
  PMLR, 2021, pp. 12\,253--12\,266.

\bibitem{NEURIPS2020_a376802c}
\BIBentryALTinterwordspacing
T.~Vogels, S.~P. Karimireddy, and M.~Jaggi, ``Practical low-rank communication
  compression in decentralized deep learning,'' in \emph{Advances in Neural
  Information Processing Systems}, H.~Larochelle, M.~Ranzato, R.~Hadsell,
  M.~Balcan, and H.~Lin, Eds., vol.~33.\hskip 1em plus 0.5em minus 0.4em\relax
  Curran Associates, Inc., 2020, pp. 14\,171--14\,181. [Online]. Available:
  \url{https://proceedings.neurips.cc/paper_files/paper/2020/file/a376802c0811f1b9088828288eb0d3f0-Paper.pdf}
\BIBentrySTDinterwordspacing

\bibitem{cheng2017survey}
Y.~Cheng, D.~Wang, P.~Zhou, and T.~Zhang, ``A survey of model compression and
  acceleration for deep neural networks,'' \emph{arXiv preprint
  arXiv:1710.09282}, 2017.

\bibitem{yang2020learning}
H.~Yang, M.~Tang, W.~Wen, F.~Yan, D.~Hu, A.~Li, H.~Li, and Y.~Chen, ``Learning
  low-rank deep neural networks via singular vector orthogonality
  regularization and singular value sparsification,'' in \emph{Proceedings of
  the IEEE/CVF conference on computer vision and pattern recognition
  workshops}, 2020, pp. 678--679.

\bibitem{attouch2013convergence}
H.~Attouch, J.~Bolte, and B.~F. Svaiter, ``Convergence of descent methods for
  semi-algebraic and tame problems: proximal algorithms, forward--backward
  splitting, and regularized gauss--seidel methods,'' \emph{Mathematical
  Programming}, vol. 137, no.~1, pp. 91--129, 2013.

\bibitem{10.1145/3298981}
\BIBentryALTinterwordspacing
Q.~Yang, Y.~Liu, T.~Chen, and Y.~Tong, ``Federated machine learning: Concept
  and applications,'' vol.~10, no.~2, Jan. 2019. [Online]. Available:
  \url{https://doi.org/10.1145/3298981}
\BIBentrySTDinterwordspacing

\bibitem{mcmahan2017communication}
B.~McMahan, E.~Moore, D.~Ramage, S.~Hampson, and B.~A. y~Arcas,
  ``Communication-efficient learning of deep networks from decentralized
  data,'' in \emph{Artificial Intel. and Statistics}.\hskip 1em plus 0.5em
  minus 0.4em\relax PMLR, 2017, pp. 1273--1282.

\bibitem{sery2020analog}
T.~Sery and K.~Cohen, ``On analog gradient descent learning over multiple
  access fading channels,'' \emph{IEEE Trans. Signal Process.}, vol.~68, pp.
  2897--2911, 2020.

\bibitem{9042352}
M.~Mohammadi~Amiri and D.~Gündüz, ``Machine learning at the wireless edge:
  Distributed stochastic gradient descent over-the-air,'' \emph{Trans. Signal
  Process.}, vol.~68, pp. 2155--2169, 2020.

\bibitem{9050465}
Y.~Du, S.~Yang, and K.~Huang, ``High-dimensional stochastic gradient
  quantization for communication-efficient edge learning,'' \emph{Trans. Signal
  Process.}, vol.~68, pp. 2128--2142, 2020.

\bibitem{yang2020federated}
K.~Yang, T.~Jiang, Y.~Shi, and Z.~Ding, ``Federated learning via over-the-air
  computation,'' \emph{IEEE Trans. Wireless Commun.}, vol.~19, no.~3, pp.
  2022--2035, 2020.

\bibitem{davenport2016overview}
M.~A. Davenport and J.~Romberg, ``An overview of low-rank matrix recovery from
  incomplete observations,'' \emph{J. Sel. Top. Signal Process.}, vol.~10,
  no.~4, pp. 608--622, 2016.

\bibitem{cai2010singular}
J.-F. Cai, E.~J. Cand{\`e}s, and Z.~Shen, ``A singular value thresholding
  algorithm for matrix completion,'' \emph{SIAM Journal on optimization},
  vol.~20, no.~4, pp. 1956--1982, 2010.

\bibitem{zhang2006schur}
F.~Zhang, \emph{The Schur complement and its applications}.\hskip 1em plus
  0.5em minus 0.4em\relax Springer Science \& Business Media, 2006, vol.~4.

\bibitem{absil2009optimization}
P.-A. Absil, R.~Mahony, and R.~Sepulchre, \emph{Optimization algorithms on
  matrix manifolds}.\hskip 1em plus 0.5em minus 0.4em\relax Princeton
  University Press, 2009.

\bibitem{luenberger2016penalty}
D.~G. Luenberger and Y.~Ye, ``Penalty and barrier methods,'' in \emph{Linear
  and Nonlinear Program.}\hskip 1em plus 0.5em minus 0.4em\relax Springer,
  2016, pp. 397--428.

\bibitem{wei2016guarantees}
K.~Wei, J.-F. Cai, T.~F. Chan, and S.~Leung, ``Guarantees of riemannian
  optimization for low rank matrix recovery,'' \emph{SIAM J. Matrix Anal. and
  Appl.}, vol.~37, no.~3, pp. 1198--1222, 2016.

\bibitem{absil2012projection}
P.-A. Absil and J.~Malick, ``Projection-like retractions on matrix manifolds,''
  \emph{SIAM J. Opt.}, vol.~22, no.~1, pp. 135--158, 2012.

\bibitem{xing2021federated}
H.~Xing, O.~Simeone, and S.~Bi, ``Federated learning over wireless
  device-to-device networks: Algorithms and convergence analysis,'' \emph{J.
  Sel. Areas Commun.}, vol.~39, no.~12, pp. 3723--3741, 2021.

\bibitem{hunger2005floating}
R.~Hunger, \emph{Floating point operations in matrix-vector calculus}.\hskip
  1em plus 0.5em minus 0.4em\relax Munich University of Technology, Inst. for
  Circuit Theory and Signal~…, 2005, vol. 2019.

\bibitem{alameddin2019toward}
S.~Alameddin, A.~Fau, D.~N{\'e}ron, P.~Ladev{\`e}ze, and U.~Nackenhorst,
  ``Toward optimality of proper generalised decomposition bases,''
  \emph{Mathematical and computational applications}, vol.~24, no.~1, p.~30,
  2019.

\bibitem{absil2015low}
P.-A. Absil and I.~V. Oseledets, ``Low-rank retractions: a survey and new
  results,'' \emph{Computational Optimization and Applications}, vol.~62,
  no.~1, pp. 5--29, 2015.

\bibitem{bergmann2019intrinsic}
R.~Bergmann and R.~Herzog, ``Intrinsic formulation of kkt conditions and
  constraint qualifications on smooth manifolds,'' \emph{SIAM J. Opt.},
  vol.~29, no.~4, pp. 2423--2444, 2019.

\bibitem{zhang2018generalized}
Z.~Zhang and M.~Sabuncu, ``Generalized cross entropy loss for training deep
  neural networks with noisy labels,'' \emph{Advances in neural information
  processing systems}, vol.~31, 2018.

\bibitem{hastie2009elements}
T.~Hastie, R.~Tibshirani, J.~H. Friedman, and J.~H. Friedman, \emph{The
  elements of statistical learning: data mining, inference, and
  prediction}.\hskip 1em plus 0.5em minus 0.4em\relax Springer, 2009, vol.~2.

\bibitem{kim2021comparing}
T.~Kim, J.~Oh, N.~Kim, S.~Cho, and S.-Y. Yun, ``Comparing kullback-leibler
  divergence and mean squared error loss in knowledge distillation,''
  \emph{arXiv preprint arXiv:2105.08919}, 2021.

\bibitem{yang2019proxsgd}
Y.~Yang, Y.~Yuan, A.~Chatzimichailidis, R.~J. van Sloun, L.~Lei, and
  S.~Chatzinotas, ``Proxsgd: Training structured neural networks under
  regularization and constraints,'' in \emph{Inter. Conf. on Learn.
  Representations}, 2019.

\bibitem{donoho2009message}
D.~L. Donoho, A.~Maleki, and A.~Montanari, ``Message-passing algorithms for
  compressed sensing,'' \emph{Proceedings of the National Academy of Sciences},
  vol. 106, no.~45, pp. 18\,914--18\,919, 2009.

\bibitem{UniformQ}
P.~Docs, ``Uniform quantization,''
  \url{https://pocketflow.github.io/uq_learner/}.

\bibitem{deng2012mnist}
L.~Deng, ``The mnist database of handwritten digit images for machine learning
  research [best of the web],'' \emph{Signal Process. Mag.}, vol.~29, no.~6,
  pp. 141--142, 2012.

\bibitem{li2021weakly}
X.~Li, S.~Chen, Z.~Deng, Q.~Qu, Z.~Zhu, and A.~Man-Cho~So, ``Weakly convex
  optimization over stiefel manifold using riemannian subgradient-type
  methods,'' \emph{SIAM J. Opt.}, vol.~31, no.~3, pp. 1605--1634, 2021.

\bibitem{liu2017convergence}
S.~Liu, Z.~Qiu, and L.~Xie, ``Convergence rate analysis of distributed
  optimization with projected subgradient algorithm,'' \emph{Automatica},
  vol.~83, pp. 162--169, 2017.

\end{thebibliography}

\end{document}